\documentclass[bg, hvmath, 11pt]{copernicus}
\usepackage[a4paper, hmargin=20mm, vmargin=20mm]{geometry}
\usepackage{wasysym,subfig}
\usepackage[pdfborder={0 0 .3}, pdfauthor={Marco van Hulten}]{hyperref}
\usepackage[all]{hypcap}
\usepackage[percent]{overpic}
\usepackage[usenames,dvipsnames]{xcolor}
\usepackage{appendix}
\usepackage[font={small}]{caption}
\usepackage{xspace}
\xspaceaddexceptions{]}
\DeclareRobustCommand{\DutchName}[4]{#2~#1}

\begin{document}\hack{\sloppy}
\def\introductionexists{true}
\def\conclusionsexists{true}

\title{On the effects of circulation, sediment resuspension and biological
incorporation by diatoms in an ocean model of aluminium}

\author[1]{M.~M.~P.~van~Hulten}
\author[1]{A.~Sterl}
\author[2,3,4]{R.~Middag}
\author[4,5]{H.~J.~W.~de~Baar}
\author[6]{M.~Gehlen}
\author[6]{J.-C.~Dutay}
\author[7]{A.~Tagliabue}

\affil[1]{Royal Netherlands Meteorological Institute (KNMI),
Utrechtseweg 297, 3731 GA De Bilt, the Netherlands}
\affil[2]{University of Otago,
364 Leith Walk, Dunedin, 9016, New Zealand}
\affil[3]{University of California Santa Cruz (UCSC),
1156 High Street, Santa Cruz, CA 95064, USA}
\affil[4]{Royal Netherlands Institute for Sea Research (NIOZ),
Landsdiep 4, 1797 SZ 't Horntje, Texel, the Netherlands}
\affil[5]{University of Groningen (RUG),
Postbus 72, 9700 AB Groningen, the Netherlands}
\affil[6]{Laboratoire des Sciences du Climat et de l'Environnement (LSCE), LSCE-Orme, point courrier 129, CEA-Orme des Merisiers, 91191 Gif-sur-Yvette Cedex, France}
\affil[7]{University of Liverpool,
4 Brownlow Street, Liverpool L69 3GP, UK}

\correspondence{M.~M.~P.~van~Hulten (marco@hulten.org)}
\runningauthor{M.M.P.~van~Hulten et~al.}

\runningtitle{An Al ocean model: circulation, sediment resuspension and incorporation}

\received{12~July 2013}
\accepted{21~July 2013}
\published{}

\maketitle

\begin{abstract}
The distribution of dissolved aluminium in the West Atlantic Ocean shows a
mirror image with that of dissolved silicic acid, hinting at intricate
interactions between the ocean cycling of Al and Si.
The marine biogeochemistry of Al is of interest because of its potential impact
on diatom opal remineralisation, hence Si availability.  Furthermore, the
dissolved Al concentration at the surface ocean has been used as a tracer for
dust input, dust being the most important source of the bio-essential trace
element iron to the ocean.
Previously, the
dissolved concentration of Al was simulated reasonably well with only a~dust
source, and scavenging by adsorption on settling biogenic debris as the only
removal process.
Here we explore the impacts of (i) a sediment source of Al in the
Northern Hemisphere (especially north of $\sim$40\degree\,N), (ii) the imposed
velocity field, and (iii) biological incorporation of Al on the modelled Al
distribution in the ocean. The sediment source clearly improves the
model results, and using a different velocity field shows the importance
of advection on the simulated Al distribution. Biological incorporation
appears to be a potentially important removal process. However,
conclusive independent data to constrain the Al\,:\,Si incorporation ratio
by growing diatoms are missing. Therefore, this 
study does not provide a definitive answer to the question of the relative
importance of Al removal by incorporation compared to removal by adsorptive
scavenging.
\enlargethispage{\baselineskip}
\end{abstract}

\begin{figure*}[ht!]
    \capstart
    \begin{overpic}[width=\textwidth,height=.3\textwidth]{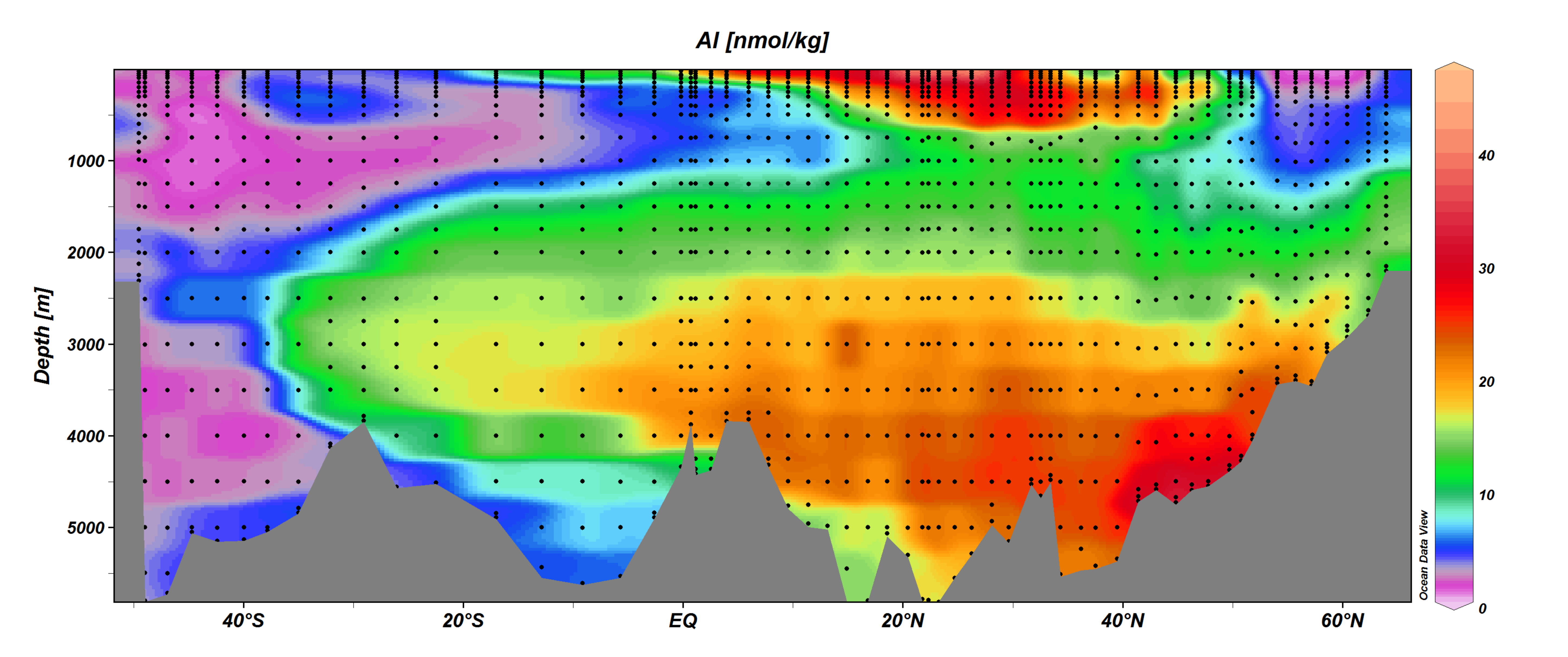}
        \put(86,6){\color{white}\bf(a)}
    \label{fig:observed_Al}
    \end{overpic}
    \begin{overpic}[width=\textwidth,height=.3\textwidth]{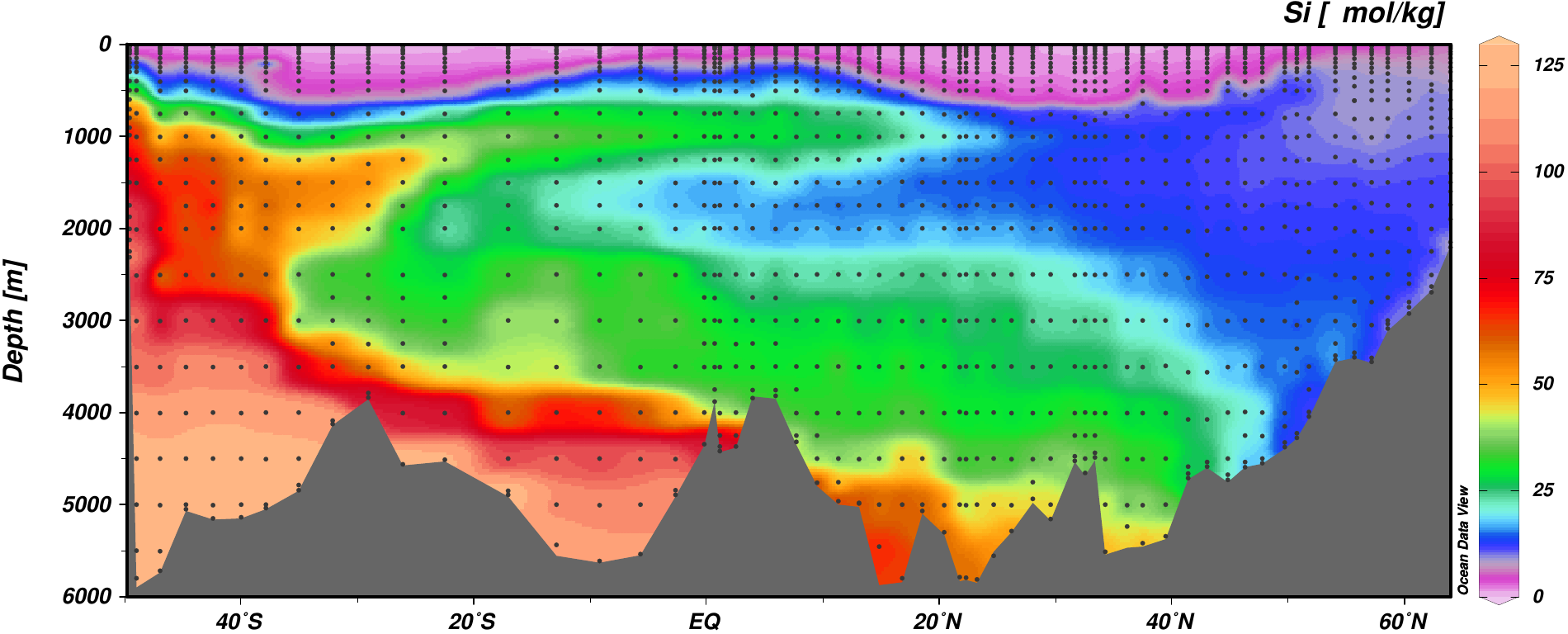}
        \put(86,6){\color{white}\bf(b)}
    \label{fig:observed_Si}
    \end{overpic}
    \caption{Observations of (a) \chem{[Al_{diss}]}\,(nM) and (b)
            \chem{[Si_{diss}]}\,(\unit{\mu M}) at the West Atlantic GEOTRACES
            transect \citep{middag:prep}.  Dots are locations of measurements.}
    \label{fig:observations}
\end{figure*}

\introduction                           \label{sec:alu_bg:introduction}
One of the first remarkable findings of the now ongoing GEOTRACES programme is
the remarkable mirror image between the distributions of
dissolved aluminium (\chem{Al_{diss}}) and dissolved silicic acid
(\chem{Si_{diss}}) in the West Atlantic Ocean (Fig.~\ref{fig:observations}),
which suggests a close interaction between
the two tracers. As \chem{Si_{diss}} is a major nutrient for
diatom growth, it is important to understand this interaction.
An initial effort towards understanding this distribution of \chem{Al_{diss}} has
been accomplished by using an ocean circulation-biogeochemistry model
\citep{vanhulten:alu_jms}.
The model only had one source, aeolian dust input at the surface,  and
one sink, scavenging by adsorption on settling biogenic debris.
This provided a reasonable agreement with the
measurements of \chem{Al_{diss}}.  However, very high \chem{Al_{diss}} near the
seafloor in the 40--50\degree\,N region was not well reproduced, which suggests
that an additional source term of Al supply from the underlying sediments in
this region is required. Furthermore, there is ample evidence and debate in the
literature on the biological incorporation of Al into the opaline
(\chem{SiO_2\cdot\mathit{n}H_2O}) frustules of growing diatoms in ocean
surface waters, and/or merely post-depositional Al enrichment of fossil opal
deposits in the sediments.  The importance of these processes needs
to be evaluated.  Finally, in the previous simulation the model
distributions of \chem{Si_{diss}} deviated significantly from the observed
distribution (Fig.~\ref{fig:observations}b), most likely due to imperfections in
the circulation of the model.

The cycling and distribution of aluminium in the ocean has received attention
for several reasons, among which are the interactions between the cycles of
aluminium (Al) and silicon (Si).  Dissolved Al is scavenged by adsorption onto
biogenic debris that settles as aggregates into the deep ocean, henceforth
called \chem{Al_{ads}} \citep[e.g.][]{stoffyn1982,orians1986,hydes1988}.
In addition, dissolved Al becomes incorporated into biological opal
(\chem{SiO_2\cdot\mathit{n}H_2O}), primarily the frustules of diatoms
\citep[e.g.][]{stoffyn1979,hydes1988,vanbeusekom1992,koning2007}.  The Al
incorporated in opal of living diatoms is hereafter called \chem{Al_{diat}}.
The relatively heavy opal (twice the density of seawater) serves as ballast for
settling aggregates; it removes the adsorbed as well as the incorporated Al
efficiently.  This is consistent with the reduced levels of dissolved Al, in
regions of high diatom production \citep[e.g.][]{orians1986}.  Conversely, the
rate of dissolution of settling opal debris (hereafter called biogenic silica,
\chem{Si_{biog}}) appears to be controlled by the Al/Si element ratio of this
opal.  The higher the Al/Si ratio of opal, the lower the rate of
\chem{Si_{biog}} dissolution
\citep[e.g.][]{lewin1961,vanbennekom1991,dixit2001,gehlen2002}.

Another major reason of interest in Al is the use of Al as a~tracer of
aeolian dust supply into the surface ocean, which is an important source of 
iron and other trace nutrients.  Indeed, it is currently assumed that the major
source of Al to the open ocean is dust deposition
\citep[e.g.][]{kramer2004,dejong2007,middag:prep}.  A~fraction of
the Al in dust (1--15\,\unit{\%}) dissolves within the upper mixed layer
\citep[e.g.][]{orians1986,maring1987,baker2006,han2012}.  Below the mixed layer
the dissolution of Al from dust is deemed negligible.  The remaining 85--99\,\%
fraction of Al remains in the particulate, lithogenic phase and sinks to the
bottom of the ocean where it is assumed to be buried in the sediment.

\begin{figure}[ht!]
    \includegraphics[width=\linewidth]{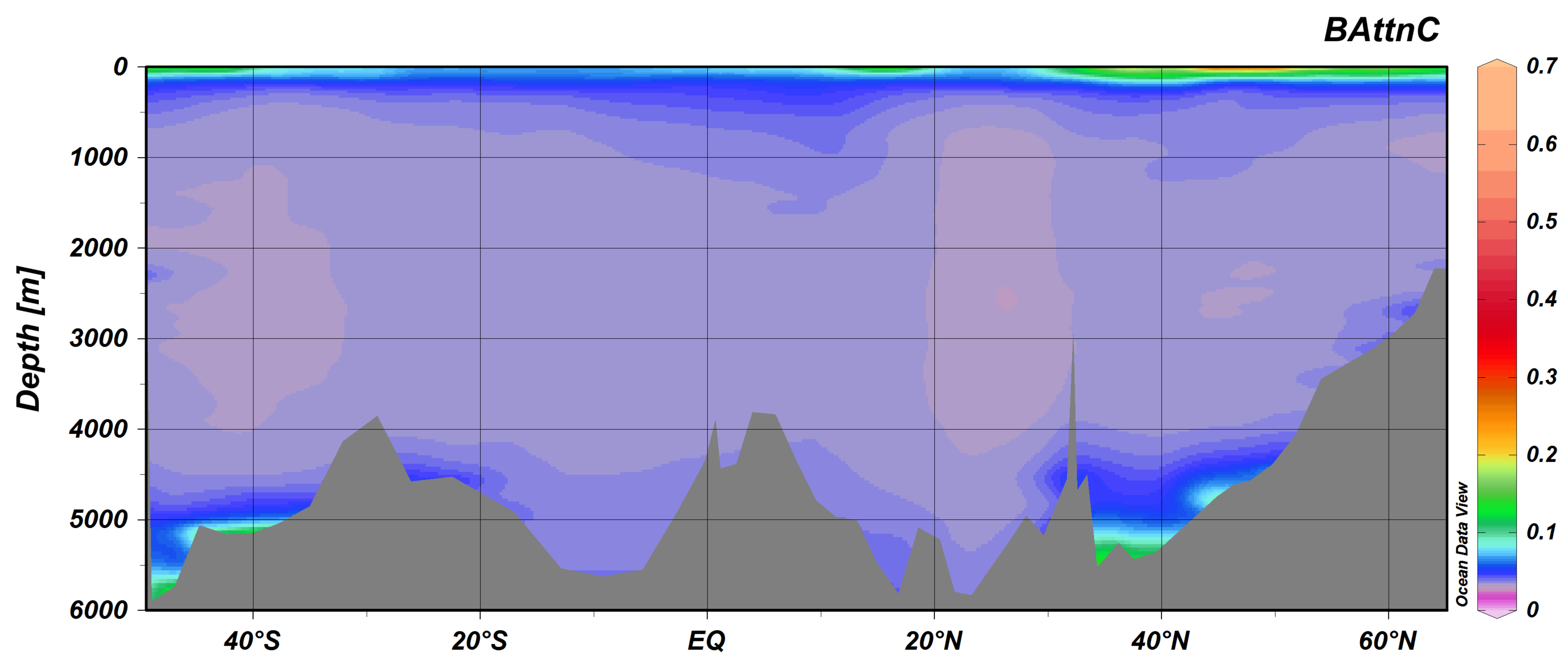}
    \caption{Beam attenuation coefficient, a measure of the amount of particles
            \citep[e.g.][]{behrenfeld2006}.  Courtesy of Micha Rijkenberg.}
    \label{fig:beam}
\end{figure}

The second source of Al is hypothesised to be sediment resuspension and
subsequent release, e.g.\ by desorption, from previously sedimented Al
\citep{moran1991,middag2012,middag:prep}.  Such a source can be contrasted with
diffusion from sediments, which typically occurs only for redox-active elements
like iron and manganese.  Indeed, a~high concentration of dissolved aluminium
(\chem{Al_{diss}}) has been measured near the deep sediment in the West Atlantic
Ocean at 45--50{\degree}\,N (Fig.~\ref{fig:observations}(a)).  One prerequisite
is sufficient turbulence near the sediment.  This is satisfied at several
locations in the West Atlantic Ocean (Fig.~\ref{fig:beam}).  Especially north of
$\sim 35${\degree}\,N and south of $\sim 40${\degree}\,S, significant
resuspension of sediment occurs \citep{biscaye1977,gross1988}.  Another obvious
prerequisite is an adequate supply of sedimenting Al towards the seafloor.  Even
though lithogenic particulate Al from dust deposition is such a~supply, Al in
that form is relatively refractory, meaning it is not easily released
\citep{brown2009}.  Hence, sedimenting Al associated with \chem{Si_{biog}} is
a~more obvious candidate for the sediment source of \chem{Al_{diss}}.  The
scavenger and incorporator of Al, \chem{Si_{biog}}, is mostly present alongside
active diatom production: north of 44{\degree}\,N and south of 40{\degree}\,S
\citep{nelson1995,treguer2013}.  Therefore, in these regions this prerequisite
is satisfied.

There appears to be variability in the sediment source of Al.  Generally,
\chem{[Al_{diss}]} is significantly higher near the sediment compared to
concentrations 500\;\unit{m} above the sediment.  However, the elevation of
[\chem{Al_{diss}}] near the sediments of the Southern Ocean is very small
compared to its elevation in the Atlantic Ocean
\citep{moran1992:weddell,middag2011:Al:Southern}.  This strongly suggests that
in the Southern Hemisphere the desorption of Al from \chem{Al_{ads}} from
resuspended sediments is very small compared to the desorption in the Atlantic
Ocean.  Upon settling on the seafloor, the desorption of adsorbed aluminium is
hypothesised to be controlled by the concentration of dissolved silicon in
ambient seawater.  At a~higher dissolved Si concentration, the desorption of Al
is reduced \citep{mackin1986}.  Especially in the Antarctic Bottom Water (AABW)
\chem{[Si_{diss}]} is very high, preventing desorption of Al.  Biogenic silica
may also play a role in preventing desorption of Al, since (adsorbed) Al is
incorporated in biogenic silica \citep{koning2007}.  Understanding these
interactions is important for understanding the cycling of Si and hence the
primary production of diatoms.

\begin{figure}
    \centering
    \includegraphics[width=\columnwidth]{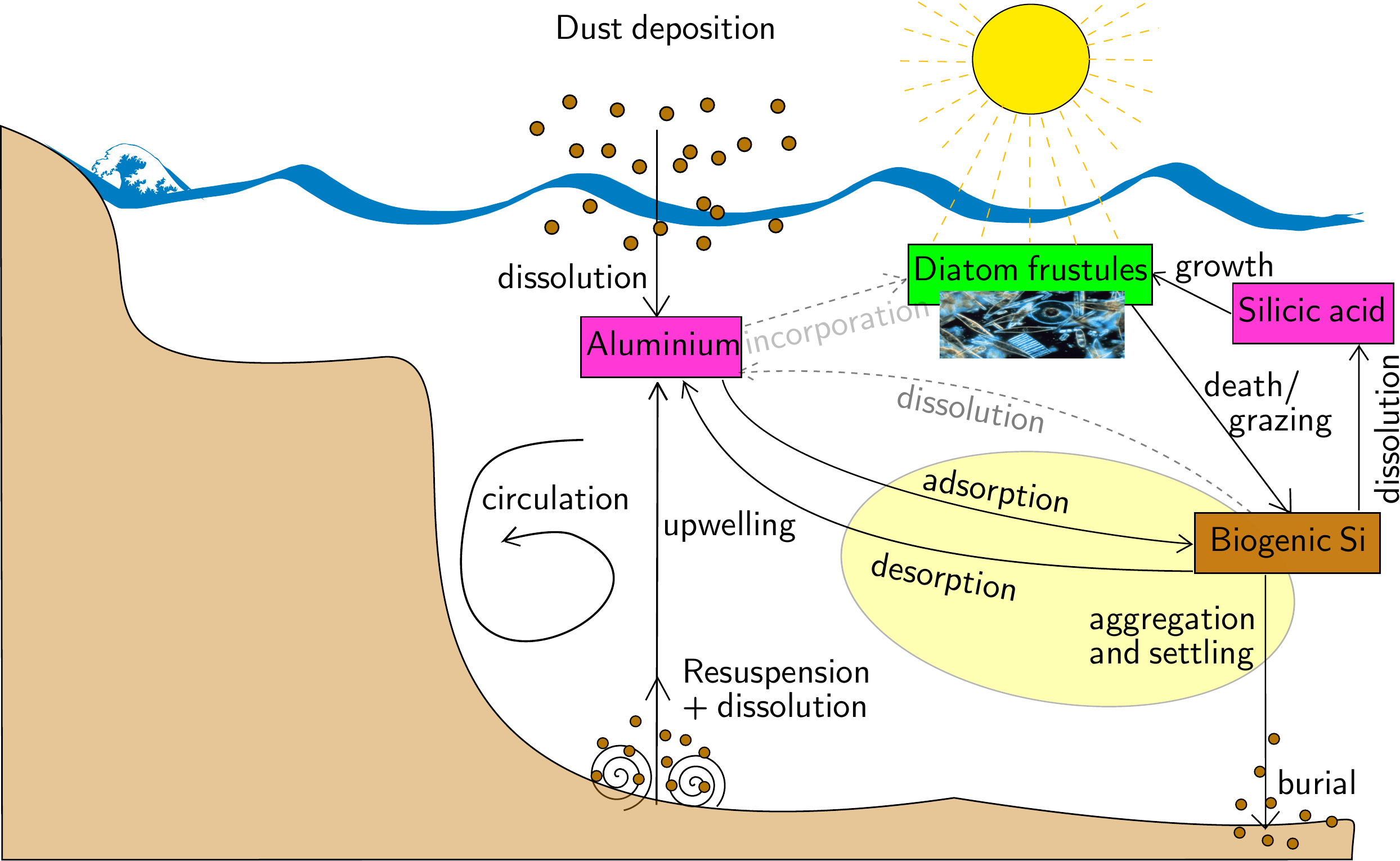}
    \caption{Aluminium cycling in the ocean.  Dissolved Al enters the ocean
        through the release of \chem{Al_{diss}} from deposited dust and
        resuspended sediments, while rivers, hydrothermal vents and reducing
        sediments are negligible sources of \chem{Al_{diss}}.  Al is mostly
        removed by \emph{reversible scavenging} (presented in the yellow
        ellipse).  The dashed arrow from aluminium to diatom frustules signifies
        incorporation of Al into diatom frustules.
        Silicic acid (\chem{Si_{diss}})
        is presented as well, since it is an essential part of understanding Al
        cycling.  However, not all sources of \chem{Si_{diss}} are presented in
        this figure.}
    \label{fig:model}
\end{figure}

Other sources do not appear to play a significant role in adding Al to the
ocean.  Even though rivers carry a large amount of Al, most of it is
removed in estuaries and continental shelf sediments, and never enters the open
ocean \citep{mackin1986control,orians1986,brown2010aluminium,jones2012}.
Finally, hydrothermal vents are not a~source of Al to the deep waters of the
oceans either \citep{hydes1986,lunel1990,elderfield1996,middag2011:Mn:Southern}.

%
The primary removal mechanism of \chem{Al_{diss}} from the surface ocean is the
adsorptive scavenging and settling with \chem{Si_{biog}} as the major carrier.
Therefore this removal is large in areas with high diatom production
\citep{stoffyn1982,orians1986,moran1988:evidence,moran1989,moran1992:kinetics}.
Besides being scavenged by surface adsorption, the \chem{Al_{diss}} becomes
incorporated as a trace substitute for Si during growth of living diatoms
\citep[e.g.][]{stoffyn1979,vanbeusekom1992,chou1997,gehlen2002,koning2007,middag2009}.
However, it is not always clear how significant the effect of incorporation is
\citep{vrieling1999,moran1988:variations,ren2011}.  Following diatom mortality,
the incorporated Al, then referred to as \emph{biogenic Al} (\chem{Al_{biog}}),
is exported with the \chem{Si_{biog}} debris.  These processes are schematically
presented in Fig.~\ref{fig:model}.

Incorporated Al is likely to inhibit the dissolution of \chem{Si_{biog}}
\citep{lewin1961,vanbennekom1991,vanbeusekom1992,dixit2001}.  This means that at
a high \chem{Al_{diat}}/\chem{Si_{diat}} ratio in living diatoms and consequent
same ratio of \chem{Al_{biog}}/\chem{Si_{biog}} in biogenic debris, less
\chem{Si_{biog}} will be remineralised.  Moreover, more silica will be buried
and hence lost from the system.  Consequently, less \chem{Si_{diss}} will be
returned to the surface through upwelling, resulting in decreased diatom
production.  This highlights the major link between Al and Si.

%
Recent years have seen the development of models of the marine biogeochemical
cycle of aluminium.  For the Al removal, \cite{gehlen2003} and
\cite{vanhulten:alu_jms} implemented a scavenging model, while \cite{han2008}
included both scavenging and biological incorporation of Al into the frustules
of diatoms.

\cite{gehlen2003} had the objective of testing the sensitivity of modelled Al
fields to dust input and thus of evaluating the possibility of constraining dust
deposition via dissolved Al near the ocean's surface.  For this purpose they
embedded an Al cycle in the HAMOCC2 biogeochemical model
\citep{maier1993geochemistry}.  The Al model consists of a~reversible
first-order relation of adsorption of \chem{Al_{diss}} onto \chem{Si_{biog}}.
In chemical equilibrium the \chem{Al_{ads}} concentration is proportional to the
product of the concentrations of \chem{Al_{diss}} and suspended particulate
\chem{Si_{biog}}.  The resulting modelled concentration of \chem{Al_{diss}} was
of the same order of magnitude as the then published observations.  The model of
\cite{vanhulten:alu_jms} used the same chemical equilibrium relation between
adsorbed and dissolved Al.  Instead of testing the effect of different dust
fields, they tested the sensitivity to the solubility of Al from dust in the
ocean surface and in the water column.  This constrained the percentage and
depth of dissolution of Al from the dust.  The coefficient partitioning
\chem{Al_{ads}} and \chem{Al_{diss}} was constrained as well with the respective
sensitivity simulation.  A sensitivity simulation with a margin sediment source
showed that margin sediments are probably not an important source of Al.  The
main goal of \cite{han2008} was to better constrain the dust deposition field.
For this purpose they used the Biogeochemical Elemental Cycling (BEC) model
improved by \cite{moore2008} as a starting point.  This was used in combination
with the Dust Entrainment And Deposition (DEAD) model to explicitly constrain
dust deposition.  In addition to scavenging, \cite{han2008} added a biological
Al uptake module where the Al\,:\,Si uptake ratio is a~function of the ambient
dissolved Al and Si concentrations.  However, they did not expand on the
importance of biological incorporation relative to adsorptive scavenging.

%
These recently developed models are consistent with the first principles of Al
cycling in the ocean, showing that the dissolution of Al from dust and the
reversible scavenging by \chem{Si_{biog}} can reproduce the main features of the
observed \chem{Al_{diss}} concentration.  However, the deep ocean has not been
simulated very well, and significant numbers of accurate deep ocean measurements
of \chem{[Al_{diss}]} have only become available recently
\citep[e.g.][]{middag2009}.  Furthermore, some studies raise doubts about the
scavenging nature of Al removal from the ocean \citep[e.g.][]{koning2007}.

Recent high-accuracy observations from the West Atlantic Ocean GEOTRACES
transect show a~mirror image between \chem{[Al_{diss}]} and \chem{[Si_{diss}]}
(Fig.~\ref{fig:observations}).  The key observation from this transect data is
that \chem{Si_{diss}} is a nutrient type (enriched in old water) and Al a
scavenged type element (depleted in older water).  Furthermore, the mirror image
suggests that (a) there is a~very modest sediment source of Al where
\chem{[Si_{diss}]} is relatively high, i.e.\ where the AABW prevails flowing
from Antarctica up to 45{\degree}\,N, and (b) the Denmark Strait Overflow Water
(DSOW) brings bottom waters with low \chem{[Si_{diss}]} from the Denmark Strait
($\sim$66{\degree}\,N) to at least 45{\degree}\,N\@.  The latter, in combination
with ample supply of adsorbed Al from opal debris of diatom blooms in overlying
surface waters, appears to result in major desorption of Al from resuspended
particles in the bottom waters \citep{middag:prep}.

The other process of interest is that of biological incorporation of Al by
diatoms.  The goal in this research is to assess whether this is a significant
process.  A simulation with incorporation is expected to yield a decrease of
\chem{[Al_{diss}]} compared to the simulation without incorporation.  The
decrease may even result in an unrealistically low \chem{[Al_{diss}]} since
scavenging parameters were tuned in the reference simulation to fit the open
ocean main thermocline distribution of \chem{Al_{diss}} and most of the West
Atlantic GEOTRACES transect at full depth.

%
In this study the model of \cite{vanhulten:alu_jms} is extended with three major
changes.  The different simulations address the role of circulation, the
importance of a sediment source and the significance of biological incorporation
of Al by diatoms.  Details on this model and of the observational dataset are
given in Sect.~\ref{sec:alu_bg:methods}.  Next, the results of the reference
simulation and the three sensitivity simulations, relative to the reference
simulation and to the simulation in \cite{vanhulten:alu_jms}, are presented in
Sect.~\ref{sec:alu_bg:results}.  More discussion about the model assumptions and
the simulation results is to be found in Sect.~\ref{sec:alu_bg:discuss}.
Finally, Sect.~\ref{sec:alu_bg:conclusion} ends with the major conclusions.

\section{Methods}               \label{sec:alu_bg:methods}
\subsection{Model description}  \label{sec:alu_bg:model}
\subsubsection{Model framework} \label{sec:alu_bg:model_framework}
In order to simulate the three-dimensional distribution of dissolved Al, the
biogeochemical model PISCES is used \citep{aumont2006,ethe2006}.  This
model has been employed for many other studies concerning trace metals, as well
as large scale ocean biogeochemistry
\citep[e.g.][]{aumont2006,gehlen2007,arsouze2009,dutay2009,tagliabue2010}.
In the simulations described here, PISCES has been driven by climatological
velocity fields obtained from the general circulation model called
\textit{Nucleus for European Modelling of the Ocean} (NEMO) \citep{madec2008} of
which the dynamical component is called \textit{Oc\'ean PArall\'elis\'e} (OPA)
\citep{madec1998}.

The model PISCES simulates the cycle of carbon, the major nutrients (nitrate,
phosphate, ammonium, silicic acid) and the trace nutrient iron, along with two
phytoplankton types (nanophytoplankton and diatoms), two zooplankton grazers
(micro- and mesozooplankton), two classes of particulate organic carbon (small
and large) of differential settling velocities, as well as calcite and biogenic
silica.  The PISCES model distinguishes three silicon pools: (i) the silicon
content of living diatoms (\chem{Si_{diat}}); (ii) the silicon content of dead,
settling diatoms (\chem{Si_{biog}}) and (iii) dissolved silicic acid
(\chem{Si_{diss}}).  In the model, \chem{Si_{diss}} and other nutrients are
supplied to the ocean by means of atmospheric dust deposition and rivers, while
iron enters the ocean as well through sediment remobilisation
\citep{aumont2006}.  The standard version of PISCES accounts for 24 tracers.
For a more detailed description of PISCES see the auxiliary material of
\cite{aumont2006}.

In this study PISCES is run off-line forced by a climatological year of monthly
physical fields including turbulent diffusion.  All model fields are defined on
the ORCA2 grid, an irregular grid covering the whole world ocean with a nominal
resolution of $2{\degree}\! \times 2{\degree}$, with the meridional
(south--north) resolution increased near the equator.  It has two ``north
poles'' in Canada and Russia to eliminate the coordinate singularity in the
Arctic Ocean.  Its vertical resolution is 10\,m in the upper 100\,m, increasing
downwards to 500\,m, such that there are 30 layers in total and the ocean has a
maximum depth of 5\,000\,m.

\subsubsection{Aluminium model}    \label{sec:alu_bg:model_aluminium}
The Al model is based on the preceding model of \citet{vanhulten:alu_jms} which
computes the concentration of dissolved aluminium (\chem{Al_{diss}}) and
adsorbed Al (\chem{Al_{ads}}).  In this study two additional tracers are
introduced to PISCES: the Al incorporated in the opaline frustules of living
diatoms (\chem{Al_{diat}}) and its biogenic debris (\chem{Al_{biog}}).  All
tracer names are listed in Table~\ref{tab:alu_bg:tracers}.  For completeness,
the Si tracers in PISCES are given as well.  The concentrations of the tracers
are indicated by square brackets [], and are given in units of
\unit{mol\,dm^{-3}} (shortly \unit{M} or molar).
\begin{table}[ht!]
\centering
\begin{tabular}{l|l}
\hline
tracer:             & description:                                      \\
\hline
\chem{Al_{diss}}    & dissolved Al: ions and colloids with $\diameter <0.2$\;\unit{\mu m} \\
\chem{Al_{ads}}     & Al adsorbed onto biogenic silica                  \\
\chem{Al_{diat}}    & Al incorporated in frustules of living diatoms    \\
\chem{Al_{biog}}    & Al incorporated in frustules of dead diatoms      \\
\hline
\chem{Si_{diss}}    & dissolved Si, or silicic acid                     \\
\chem{Si_{diat}}    & Si incorporated in frustules of living diatoms    \\
\chem{Si_{biog}}    & Si incorporated in frustules of dead diatoms      \\
\hline
\end{tabular}
\caption{Aluminium and silicon tracers in the model.}
\label{tab:alu_bg:tracers}
\end{table}

There are two different sources of Al in the model.  One source is via the
dissolution of dust particles in the upper ocean layer.  The dust deposition
field was taken from the output of the atmospheric dust model INCA
\citep{hauglustaine2004,textor2006}.  The other source, which has now been added
to the previous model, is sediment resuspension and subsequent dissolution.
When \chem{Al_{ads}} reaches the ocean floor, it is assumed to be buried, except
for the resuspension and subsequent dissolution of Al in some simulations.  The
model is schematically represented in Fig.~\ref{fig:model}, where the extensions
to \citet{vanhulten:alu_jms} are resuspension/desorption (bottom-left) and
biological incorporation (top-right).  Sediment and porewater reactions are not
explicitly modelled. Resuspension and subsequent desorption take place in the
water column.

The model parameters are listed in Table~\ref{tab:parameters}.
\begin{table}[ht!]
\begin{tabular}{l|l|l}
\tophline
Parameter:                  & Symbol:            & Value:                   \\
\middlehline
mass fraction of Al in dust & $f_\text{Al}$      & 8.1\,{\%}                \\
surface dissolution fraction& $\alpha_\text{Al}$ & 5\,{\%}                  \\
partition coefficient       & $k_d$              & $112\cdot 10^3$\;\unit{dm^3\,mol^{-1}} \\
first-order rate constant   & $\kappa$           & $10^4$\;\unit{yr^{-1}}   \\
settling velocity of \chem{Al_{ads}} and \chem{Al_{biog}} & $w_s$ & 30--200\;\unit{m\,d^{-1}} \\
\bottomhline
\end{tabular}
\vskip 2mm
\caption{Parameters for the reference simulation RefDyn2 (identical to those of
RefDyn1.}
\label{tab:parameters}
\end{table}
The aluminium fraction in dust ($f_\text{Al}$) is based on the mass percentages
of Al known to be present in the Earth's crust.  This is about 8.1\,\unit{\%_m}
aluminium on average \citep{wedepohl1995}.  Most of this Al consists of oxides
that do not dissolve easily.  The fraction of Al from dust ($\alpha_\text{Al}$)
that dissolves is not well constrained but is probably in the range of
1--15\,{\%} \citep{orians1986,jickells2005}.  Here $\alpha_\text{Al} = 5\,{\%}$
is chosen, since this is has been successfully used in previous modelling work
\citep{han2008,vanhulten:alu_jms}.  The dissolution occurs only in the upper
model layer (0--10\;\unit{m} depth range), and is described by the following
equation:
\begin{equation}
    \frac{\partial \chem{[Al_{diss}]}}{\partial t}\Big|_\text{deposition} = \frac{\alpha_\text{Al}
        \cdot f_\text{Al}}{m_\text{Al} \cdot \Delta z_1} \cdot \mathit{\Phi}_\text{dust} \;,
\end{equation}
where $m_\text{Al}$ is the atomic mass of Al, $\Delta z_1 = 10$\;\unit{m} is the
thickness of the surface model layer and $\mathit{\Phi}_\text{dust}$ is the dust
flux into the ocean.  The Al that does not dissolve from dust is assumed to play
no role in the biogeochemical cycle of Al on our timescales of interest, and can
be thought of as being buried in marine sediments.

Another source of Al is sediment resuspension (bottom-left in
Fig.~\ref{fig:model}).  In reality, this is induced by near-sediment turbulence,
creating a 200 to 1000\,m thick nepheloid layer above the sediment containing
significant amounts of suspended sediment particles
\citep[e.g.][]{lampitt1985,hwang2010}.  However, here it is assumed that
recently settled \chem{Al_{ads}} is resuspended and subsequently partly
dissolved in the bottom model layer.  Since the resuspension depends on settling
of \chem{Al_{ads}}, the relevant model equation will be introduced at the end of
this section, after the settling equation.

%
Dissolved Al is assumed to adsorb onto biogenic silica particles, while other
particles do not have an effect on the removal of \chem{Al_{diss}} (discussed in
\citet{vanhulten:alu_jms} and Sect.~\ref{sec:alu_bg:discuss}).  Hence, aside
from external inputs (and optionally the incorporation within the silica), the
\chem{Al_{diss}} concentration is governed by adsorption and desorption (yellow
ellipse in Fig.~\ref{fig:model}).  The \chem{Al_{diss}} and \chem{Al_{ads}}
concentrations are governed by the following reversible first-order adsorption
equation \citep{vanhulten:alu_jms}:
\begin{equation}
    \frac{\partial \chem{[Al_{ads}]}}{\partial t} \Big|_\text{ad/desorption}
        = \kappa (A_\text{ads}^\text{eq} - \chem{[Al_{ads}]}) \;,
    \label{eqn:scav}
\end{equation}
where
\begin{equation}
  A_\text{ads}^\textrm{eq} = k_d \cdot \chem{[Al_{diss}]} \cdot \chem{[Si_{biog}]} \;,
  \label{eqn:equil}
\end{equation}
in which $A_\text{ads}^\textrm{eq}$ is the chemical equilibrium concentration of
\chem{Al_{ads}}.  The parameter $k_d$ (\unit{dm^3\,mol^{-1}}) is the partition
coefficient and $\kappa$ (\unit{s^{-1}}) is the first-order rate constant for
equilibration of \chem{[Al_{ads}]} to $A_\text{ads}^\text{eq}$\@.  Finally,
\chem{Si_{biog}} is the biogenic silica concentration, as with all
concentrations, in \unit{mol\,dm^{-3}}.  Since total Al is conserved when only
internal processes are concerned, the time derivative of \chem{[Al_{diss}]}
equals the negative of that of \chem{[Al_{ads}]}.

%
As an extension to the original model, aluminium is incorporated into the
frustules of diatoms during production.  The diatom incorporation of Al is
modelled by multiplying the rate of production of diatom opal (\chem{Si_{diat}})
with the dissolved Al/Si concentration ratio in ambient seawater, with some
refinements as explained below.  For the biological Si cycle, production and
mortality (including grazing by micro- and mesozooplankton) of diatoms, and
dissolution of debris \chem{Si_{biog}}, are represented by prod, mort and diss,
respectively, in Eqs~\eqref{eqn:incorp}.  All rate variables are proportional to
the ratio of Al and Si in the relevant source pool.  Accordingly, the biological
model equations for Al are as follows:
\begin{subequations}
\begin{align}
    \frac{\partial \chem{[Al_{diat}]}}{\partial t} \Big|_\text{bio} &= R_\text{Al:Si} \cdot \text{prod}
                            - \frac{\chem{[Al_{diat}]}}{\chem{[Si_{diat}]}} \cdot \text{mort}
    \label{eqn:diat} \\
    \frac{\partial \chem{[Al_{biog}]}}{\partial t} \Big|_\text{bio} &= \frac{\chem{[Al_{diat}]}}{\chem{[Si_{diat}]}} \cdot \text{mort}
                            - \frac{\chem{[Al_{biog}]}}{\chem{[Si_{biog}]}} \cdot \text{diss}
    \label{eqn:biog} \\
    \frac{\partial \chem{[Al_{diss}]}}{\partial t} \Big|_\text{bio} &= \frac{\chem{[Al_{biog}]}}{\chem{[Si_{biog}]}} \cdot \text{diss}
                            - R_\text{Al:Si} \cdot \text{prod}   \;.
    \label{eqn:diss}
\end{align}
\label{eqn:incorp}%
\end{subequations}
No further refinement is made in any process, except, optionally, for
incorporation during production.  This is accomplished by the incorporation
ratio $R_\text{Al:Si}$, which is defined as:
\begin{equation}
    R_\text{Al:Si} = \text{min} \left( c_\text{in} \cdot
        \frac{\chem{[Al_{diss}]}}{\chem{[Si_{diss}]}},\; r_\text{max} \right)\;,
    \label{eqn:xi}
\end{equation}
where \chem{[Al_{diss}]/[Si_{diss}]} is the ratio of dissolved Al and Si, and
$c_\text{in}$ ($0\leq c_\text{in}\lesssim 1$) is an optional weight factor for
the \chem{Al}:\chem{Si} incorporation ratio, and $r_\text{max} \geq 0$ is an
optional prescribed maximum value for the \chem{Al_{diat}}/\chem{Si_{diat}}
ratio within the opal of living diatoms.  These two optional parameters are not
used in the model simulations presented in this work (i.e.\ we use
$c_\text{in}=1$ and $r_\text{max}=\infty$), but will be discussed in
Sect.~\ref{sec:alu_bg:discuss_biological} on biological incorporation.  The
biological extension to the model is schematically presented in
Fig.~\ref{fig:model} as the dotted line from \emph{Aluminium} to \emph{Diatom
frustules}.

%
Both \chem{Al_{biog}} and \chem{Al_{ads}} settle through the water column,
because the concurrent biogenic silica is denser than seawater.  Adsorbed and
incorporated Al (their concentrations both denoted as $A_\text{part}$) settle
along with biogenic silica with a~velocity $w_s$, varying from
30\;\unit{m\,d^{-1}} in the upper 100\;\unit{m}, to 200\;\unit{m\,d^{-1}} at
4\;\unit{km} below the mixed layer, according to
\begin{equation}
    \frac{\partial A_\text{part}}{\partial t} \Big|_\text{settling}
        = - w_s \frac{\partial A_\text{part}}{\partial z} \;.
    \label{eqn:sink}
\end{equation}
While settling through the water column, \chem{Al_{biog}} and \chem{Al_{ads}}
may underway remineralise (Eqs~\eqref{eqn:biog} and \eqref{eqn:scav},
respectively), so adsorption and incorporation do not mean that all Al is
removed immediately from the model domain.  Part of the Al dissolves and may
upwell, and may become incorporated or scavenged once again at a~later time.

Burial of \chem{Al_{ads}} and \chem{Al_{biog}} proceeds in the bottom model
layer immediately above the sediment according to
\begin{equation}
    \partial A_\text{part} / \partial t = -(w_s\cdot
        A_\text{part}) / \Delta z_\text{bottom} \;,
    \label{eqn:burial}
\end{equation}
where $\Delta z_\text{bottom} \leq 500$\,m is the thickness of the bottom layer.
It is of the same order of magnitude as the real resuspension, or
nepheloid, layer.  In two of the simulations, part of the sedimented
\chem{Al_{ads}} is resuspended and dissolved in the bottom water layer according
to:
\begin{equation}
    \frac{\partial\chem{[Al_{diss}]}}{\partial t} \Big|_\text{resusp} = 
            \beta \cdot \frac{w_s \cdot \chem{[Al_{ads}]}}{\Delta z_\text{bottom}} \;,
    \label{eqn:resusp}
\end{equation}
where $\beta$ is a~constant ($0 < \beta \leq 1$), or some scalar function to be
defined later (Eq.~\eqref{eqn:beta_Mackin}), representing the fraction of
resuspended and subsequently dissolved Al.  Note that in the model only adsorbed
Al is redissolved, while biogenic Al is not (see
Sect.~\ref{sec:alu_bg:discuss_resusp_short} for rationale)  This process is
different from the sediment source in \citet{vanhulten:alu_jms}.  In that study
sediment input was modelled analogous to the diffusive iron source, while in
this study the redissolution depends on freshly sedimented \chem{Al_{ads}}.  The
underlying rationale and the derivation of Eq.~(\ref{eqn:resusp}) are given in
Appendix~\ref{sec:alu_bg:derivation}.  Moreover, silicic acid near the sediment
apparently inhibits the dissolution of \chem{Al_{ads}}.  This is probably
because \chem{Al_{diss}} and \chem{Si_{diss}} are stoichiometrically saturated
\citep{mackin1986}.  Observations suggest a significant Al sediment source below
Denmark Strait Overflow Water (DSOW) where \chem{Si_{diss}} concentrations are
below 30\;\unit{\mu M} (1\;\unit{\mu M}\;=\;$10^{-6}\;\unit{mol\,dm^{-3}}$),
while they suggest only a~very weak source from the sediment below AABW, which
has much higher \chem{[Si_{diss}]}, exceeding 50\;\unit{\mu M}
\citep{middag:prep}.  In one of the simulations, the bottom water
\chem{[Si_{diss}]} will be used as an inhibitor of the redissolution of Al from
resuspended sediment.

The source code of the model is included as an electronic supplement.  This code
can be used with NEMO-PISCES (in this study version 3.1 was used), to be found
at \url{http://www.nemo-ocean.eu/} (libre software licensed under
\href{http://www.cecill.info/}{CeCILL}).

\subsubsection{Simulations}             \label{sec:alu_bg:model_simulations}
%
The reference simulation (RefDyn2) is based on the ``reference experiment'' from
\citet{vanhulten:alu_jms} (RefDyn1), but there is one notable improvement.  A
different set of dynamical fields (advection, turbulent diffusion and mixing) is
now used.  The velocity fields used by \citet{vanhulten:alu_jms} (Dynamics~1)
had an Atlantic Meridional Overturning Circulation (MOC) that is too shallow and
too weak, while the northward flux of AABW is too strong, compared to estimates
based on measurements.  The fields used in this study, henceforth referred to as
Dynamics~2, are a climatology from a NEMO-OPA simulation forced by the DFS3
air-sea fluxes \citep{large2004,uppala2005,brodeau2010}.  Dynamics~2 has a more
realistic Atlantic overturning, compared to Dynamics~1.  Firstly, the depth of
North Atlantic Deep Water is about 2\,km in \citet{vanhulten:alu_jms} while it
is closer to the more reasonable 3\,km in the dynamics used here.  Secondly, the
maximum of the Atlantic Overturning Stream Function (OSF) in the AABW dropped
from about 6--7\,Sv in Dynamics~1 to about 2--3\,Sv in Dynamics~2, much more
comparable to estimates based on observations \citep{talley2003,rayner2011}.
This results in a significantly better simulation of the \chem{Si_{diss}}
distribution (see discussion in Sect.~\ref{sec:alu_bg:discuss_underlying}).  The
\chem{Si_{diss}} concentration in AABW is much closer to the observational
concentration, compared with the dynamical fields used by
\citet{vanhulten:alu_jms} (basic statistics for the West Atlantic Ocean are
presented in Sect.~\ref{sec:alu_bg:dynamics}).  This will especially be
important for one of the sediment resuspension simulations described below,
since that simulation depends on \chem{[Si_{diss}]} in bottom water.

The reference simulation (RefDyn2) is initialised from the steady state
concentrations of the reference experiment in \citet{vanhulten:alu_jms}
(RefDyn1), and is then run for another 1750 model years to steady state with
Dynamics~2.  The resulting total Al budget (both dissolved and adsorbed) in the
world ocean in the reference simulation is around 6 Tmol (1\;Tmol =
$10^{12}$\;mol).  After 1000\;\unit{yr} of simulation the Al budget (total
\chem{Al_{diss}} + \chem{Al_{ads}} integrated over the ocean volume) does not
change much (less than 1\,{\%} before reaching steady state).  Therefore, from
year 1000 our two sets of sensitivity simulations with sediment resuspension and
biological incorporation are forked and run for another 500\;\unit{yr},
alongside the reference simulation, to quasi-steady-state.  An overview of the
simulations with the key parameters is given in Table~\ref{tab:simulations}, and
the two sets of sensitivity simulations are defined and explained below.
\begin{table}[h!]
\begin{tabular}{l|l|ll|ll}
\tophline
Simulation:  & dyn & margin  & \chem{Si_{diss}} & incorp & $k_d$ \\
\middlehline
RefDyn1       & 1       & no        & --            & no            & norm  \\
RefDyn2       & 2       & no        & --            & no            & norm  \\
SedProp       & 2       & yes       & no            & no            & norm  \\
SedMackin     & 2       & yes       & yes           & no            & norm  \\
Incorp        & 2       & no        & --            & yes           & norm  \\
IncorpLowScav & 2       & no        & --            & yes           & norm/4\\
\bottomhline
\end{tabular}
\caption{Overview of the two reference simulations as defined in the above main
text, and the two sets of sensitivity simulations as defined in the below main
text.  If not specified otherwise, when the term \emph{reference simulation} is
used, we refer to RefDyn2.  Here RefDyn1 refers to the old ``reference
experiment'' of \citet{vanhulten:alu_jms}.  Dynamics~1 refers to the climatology
based on ERS satellite data; Dynamics~2 is based on DFS3.  The
\chem{Si_{diss}}-dependence refers to the dependence on bottom water
[\chem{Si_{diss}}] for dissolution of resuspended adsorbed Al.  The normal
partition coefficient is $k_d = 112\cdot 10^3$ dm$^3$\,mol$^{-1}$.  This is
equal to the $k_d = 4\cdot 10^6$\;\unit{dm^3\,kg^{-1}} in
\citet{vanhulten:alu_jms}, where the biogenic Si concentration was denoted in
\unit{kg\,dm^{-3}} instead of \unit{mol\,dm^{-3}}. For consistency in the
concentration units, we changed the unit (and hence the value) of $k_d$.}
\label{tab:simulations}
\end{table}

The dust dissolution and scavenging parameters as used by
\citet{vanhulten:alu_jms} resulted in a good simulation of \chem{[Al_{diss}]} in
the upper ocean.  Therefore, the same parameters were used for the new reference
simulation (RefDyn2, Table~\ref{tab:parameters}).  This is also the case for
most sensitivity simulations, except for IncorpLowScav where the partition
coefficient $k_d$ is decreased.  A sediment source is included in simulations
SedProp and SedMackin, and biological incorporation of Al is present in the
simulations Incorp and IncorpLowScav (Table~\ref{tab:simulations}).

%
In the first of our two sediment resuspension simulations, 60\,{\%} of
sedimented \chem{Al_{ads}} is redissolved just above the sediment ($\beta =
0.60$, see Eq.~\eqref{eqn:resusp}).  This means that redissolution is
proportional to sedimentation, hence this simulation is named SedProp.  However,
the West Atlantic Ocean GEOTRACES observations show a large elevation of
\chem{[Al_{diss}]} due to resuspension only in the Northern Hemisphere and only
very little in the Southern Hemisphere (Fig.~\ref{fig:observations}(a)).  It
appears that the process of release of resuspended sediments is inhibited in the
Southern Hemisphere.  As shown by \cite{mackin1986}, the inhibition may be
caused by high \chem{[Si_{diss}]}, which is very high in the AABW in the
southern hemisphere and flows as far north as $\sim$40{\degree}\,N, albeit
somewhat diluted by vertical mixing with overlying NADW that has lower
\chem{Si_{diss}} concentrations.  Therefore the result of \cite{mackin1986} is
used to arrive at the following dissolution fraction:
\begin{equation}
    \beta = \beta_0 \cdot (\chem{[Si_{diss}]}_\text{bottom}/\unit{\mu M})^{-0.828} \;,
    \label{eqn:beta_Mackin}
\end{equation}
where $\beta_0$ is a dimensionless constant (in this simulation set to 16.85)
and the $-0.828$ is from \cite{mackin1986}.  For further discussion see the
derivation in Appendix~\ref{sec:alu_bg:derivation} and the discussion in
Sect.~\ref{sec:alu_bg:discuss_resusp}.  Henceforth this simulation is referred
to as SedMackin, since $\beta$ is a smooth function of \chem{[Si_{diss}]}.

%
Finally, one set of two simulations with biological incorporation of Al into
diatom frustules has been performed.  In these simulations \chem{Al_{diss}} is
scavenged in the same way as in the other simulations.  No limitation to the
incorporation of \chem{Al_{diss}} has been applied.  This means that for these
simulations, $c_\text{in}=1$ and $r_\text{max}=\infty$ (see Eq.~\eqref{eqn:xi}),
such that $R_\text{Al:Si} = \chem{[Al_{diss}]}/\chem{[Si_{diss}]}$; alternatives
will be discussed in Sect.~\ref{sec:alu_bg:results}.  The first of two
incorporation simulations, Incorp, is with the normal scavenging parameters.
The second one, IncorpLowScav, is different with respect to Incorp in that the
partition coefficient is decreased from 112 to $28\cdot
10^3$\;\unit{dm^3\,mol^{-1}}, for reasons that will become clear in
Sect.~\ref{sec:alu_bg:results}.

\subsection{Observational datasets}     \label{sec:alu_bg:observations}
Recent \href{http://www.geotraces.org/}{GEOTRACES} observations of
\chem{[Al_{diss}]} in the Arctic Ocean \citep{middag2009}, North-east Atlantic
Ocean \citep[][Chapter 5]{middag2010phd}, West Atlantic Ocean
\citep{middag:prep}, the Atlantic sector of the Southern Ocean
\citep{middag2011:Al:Southern,middag2012,middag2013:Weddell} and the region
south of Australia \citep{remenyi2013phd} are used for a detailed comparison and
optimisation of the model parameters.  See the upper part of Table
\ref{tab:data} for these datasets.  These datasets comprise overall 4013
individual data values for dissolved Al. All values have been verified versus
international reference samples and their consensus values of the SAFe and
GEOTRACES programmes.  See \citet{vanhulten:alu_jms} for more details.  In
addition to the datasets used by \citet{vanhulten:alu_jms}, the GEOTRACES
International Polar Year (GIPY) data by \cite{remenyi2013phd} (data and
dissertation on request available from lead author) have been added to the
dataset, namely SAZ-Sense (GIPY2) and SR3 (GIPY6).  Those data are located
between Tasmania and Antarctica (438 data points).  Furthermore, there are five
additional stations (120 extra data points) in the North-west Atlantic Ocean
(64\,PE\,358).  These additional cruises add up to 558 extra data points,
compared to the number used by \citet{vanhulten:alu_jms}.

For a worldwide global ocean comparison one also has to rely on data that was
collected in the era before the reference samples of SAFe and GEOTRACES were
available. Inevitably, the criteria for selecting such previously published
datasets are less strict; see \citet{vanhulten:alu_jms} and
\citet[][pp.~216--8]{middag2010phd} for the criteria used for each of the
selected datasets.  The selected pre-GEOTRACES datasets are listed in the lower
part of Table~\ref{tab:data}.
\begin{table}
\begin{tabular}{l|l|l|r}
\tophline
Cruise      &   Ocean           &   Source                      & \#   \\
\middlehline
ARK XXII/2  &   Arctic          & \cite{middag2009}             & 1080 \\
ANT XXIV/3  &   Southern        & \cite{middag2011:Al:Southern} &  919 \\
SAZ-Sense   &   Southern        & \cite{remenyi2013phd}         &  146 \\
SR3         &   Southern        & \cite{remenyi2013phd}         &  292 \\
64 PE 267   &   Atlantic        & \cite{middag:prep}            &  137 \\
64 PE 319   &   Atlantic        & \cite{middag:prep}            &  383 \\
64 PE 321   &   Atlantic        & \cite{middag:prep}            &  504 \\
64 PE 358   &   Atlantic        & \cite{middag:prep}            &  120 \\
JC057       &   Atlantic        & \cite{middag:prep}            &  432 \\
\middlehline
\multicolumn{3}{l|}{Subtotal used primarily for detailed comparison} & 4013 \\
\middlehline
\multicolumn{4}{l}{}    \\
\middlehline
IOC96       & Atlantic          & \cite{vink2001}               & 1048 \\
M\;60       & Atlantic          & \cite{kremling1985}           &   91 \\
IRONAGES\!  & Atlantic          & \cite{kramer2004}             &  181 \\
EUCFe       & Pacific           & \cite{slemons2010}            &  195 \\
MC-80       & Pacific           & \cite{orians1986}             &   92 \\
VERTEX-4    & Pacific           & \cite{orians1986}             &   54 \\
VERTEX-5    & Pacific           & \cite{orians1986}             &   59 \\
KH-98-3     & Indian            & \cite{obata2007}              &  152 \\
\middlehline
\multicolumn{3}{l|}{Subtotal of other observations for global comparison} & 1872 \\
\middlehline
\middlehline
\multicolumn{3}{l|}{Grand total of all dissolved Al values} & 5885 \\
\bottomhline
\end{tabular}
\caption{Observational data used for comparison with model.}
\label{tab:data}
\end{table}

\subsection{Data--model comparison}     \label{sec:alu_bg:comparison}
For the, mostly qualitative, visual comparison between model and observations,
horizontal and vertical cross sections of the model data are plotted.  For the
horizontal \chem{[Al_{diss}]} sections four different depths are presented,
where `surface' signifies the average over the upper 30\;\unit{m},
`500\;\unit{m}' is 450--550\;\unit{m} averaged, `2500\;\unit{m}' is
2450--2550\;\unit{m} averaged and `4500\;\unit{m}' is 4450--4550\;\unit{m}
averaged.  The respective observations (same depth ranges) are presented as
coloured dots.  The colour scale is not linear to better show the main features
at both low and high concentrations of \chem{Al_{diss}}.

The vertical \chem{[Al_{diss}]} sections are of the GEOTRACES cruises in the
West Atlantic Ocean (64\;PE\,267, 319, 321, 358; and JC057 from
Table~\ref{tab:data}) and the Zero Meridian Southern Ocean (part of
ANT\;XXIV/3).  These sections are calculated from the three-dimensional model
data by converting the ORCA2-gridded model data to a rectilinear mapping, and
interpolating the rectilinear data onto the cruise track coordinates.  One of
these figures also shows the Atlantic Overturning Stream Function (OSF), defined
as the zonally (through the Atlantic Ocean) and vertically (from the surface
downward) integrated meridional current speed.  This is used as a measure for
the MOC\@.

The focus of this study is the West Atlantic Ocean for several reasons.
Firstly, recently measurements have been carried out in that region, resulting
in a large consistent (one method) dataset.  Secondly, there are too few
high-quality observations in other regions of the ocean, making it very
difficult to define a reasonable goodness of fit.  Thirdly, the West Atlantic
Ocean is of large importance to the MOC and hence the deep ocean cycling of
nutrients.  For these reasons all quantitative arguments in this study concern
the West Atlantic GEOTRACES transect.

\subsubsection{Statistics}              \label{sec:alu_bg:statistics}
To compare quantitatively the model results with the observations, we focus only
on the 1576 data points of the West Atlantic Ocean GEOTRACES transect
(Sect.~\ref{sec:alu_bg:results}).  First the observations are linearly
interpolated onto the model grid.  Then several statistics are determined,
namely the Root Mean Square Deviation (RMSD), the Reliability Index (RI) and the
correlation coefficient $r^2$.  These statistics are all based on
\cite{stow2009}, but the first two are adjusted for the inhomogeneous sample
distribution in depth.

The RMSD is determined by the following equation:
\begin{equation}
  D_{\updownarrow} = \sqrt{ \frac{ \sum_{k=1}^{30}{\Delta z_k \cdot \sum_{j=1}^{60}(m_{jk} -
o_{jk})^2} } { 60\cdot \sum_{k=1}^{30}{ \Delta z_k } } }  \;,
\end{equation}
where $o$ is the observed and $m$ the modelled \chem{[Al_{diss}]}, weighting
with model layer thickness $\Delta z_k$ of layer $k \in \{1..30\}$ for every
station $j \in \{1..60\}$.  The $\updownarrow$ signifies the vertical weighting
modification of the standard RMSD.  This is done to compensate for the
overrepresentation of data points near the ocean surface.  Furthermore, for each
sensitivity simulation we calculate the significance of the change in the RMSD
compared with the corresponding reference simulation.  This is determined by
means of a Monte Carlo simulation on the reference simulation for which a
subsample of 400 has been randomly selected from the original set of 1800
data--model points.  They are the pairs of observations and model data, both on
the model grid.  This is done 50\,000 times and from this the $2 \sigma$
confidence interval is calculated (the mean $\pm$ two times the standard
deviation).  Suppose that we wish to simulate $q$, and assume $q$ is in steady
state.  For each model $Y$ resulting in $q_Y(\mathbf{x})$, the average RMSD of
the Monte Carlo simulation of $q_Y(\mathbf{x})$ must be outside the $2 \sigma$
confidence range of the RMSD distribution of $q_X(\mathbf{x})$ to say that $Y$
is a significant improvement or worsening compared to $X$.

The reliability index adjusted by weighting with the model layer thickness is
defined as:
\begin{equation}
    RI_\updownarrow = \text{exp}\sqrt{ \frac{\sum\limits_{k=1}^{30} \Delta z_k
\sum\limits_{j=1}^{60} (\text{log}\frac{o_{jk}}{m_{jk}})^2 }{60\cdot\sum_{k=1}^{30}\Delta z_k}} \;.
\end{equation}

Finally, the correlation coefficient is defined as:
\begin{equation}
    r = \frac{\sum\nolimits_{i=1}^{1800} (o_i-\overline{o})(m_i-\overline{m})}{\sqrt{
        \sum\nolimits_{i=1}^{1800}(o_i-\overline{o})^2 \sum\nolimits_{i=1}^{1800}(m_i-\overline{m})^2}} \;,
\end{equation}
where 1800 is the total number of measurements of the West Atlantic GEOTRACES
transect and the bars denote averages.

\section{Results}                       \label{sec:alu_bg:results}
In this section the results of the simulations are presented.  The relevant
tracers of the raw model output can be found at
\url{http://data.zkonet.nl/index.php?page=Project_view&id=2916&tab=Datasets}.

\begin{table*}
\begin{tabular}{l|l|lll|l|l|l}
\tophline
Simulation:         & $D$  & $D_{\updownarrow}$& \textit{significance}$_\updownarrow$ & $2\sigma$ \textit{range}$_\updownarrow$ & $r$ & $RI$ & $RI_\updownarrow$  \\
\bottomhline
\tophline
\multicolumn{8}{l}{\chem{[Si_{diss}]}:} \\
\middlehline
RefDyn1 \citep{vanhulten:alu_jms} & 23.8\;\unit{\mu M} & 39.7\;\unit{\mu M} & -           & [34.8, 44.4] & 0.70 & 4.05  & 2.57  \\
RefDyn2 (this study)              & 12.7\;\unit{\mu M} & 18.0\;\unit{\mu M} & improvement & [15.8, 20.2] & 0.77 & 4.02  & 2.07  \\
\bottomhline
\tophline
\multicolumn{8}{l}{\chem{[Al_{diss}]}:} \\
\middlehline
RefDyn1             & 7.4\;\unit{nM}       & 8.3\;\unit{nM}       & -                 & [7.5, 9.2]   & 0.71     & 2.00  & 1.91  \\
RefDyn2             & 8.9\;\unit{nM}       & 8.7\;\unit{nM}       & insignificant     & [7.9, 9.6]   & 0.71     & 2.10  & 1.98  \\
\middlehline
RefDyn2             & 8.9\;\unit{nM}       & 8.7\;\unit{nM}       & -                 & [7.9, 9.6]   & 0.71     & 2.10  & 1.98  \\
SedProp             & 7.0\;\unit{nM}       & 5.2\;\unit{nM}       & improvement       & [4.6, 5.8]   & 0.75     & 1.97  & 1.68  \\
SedMackin           & 7.1\;\unit{nM}       & 5.9\;\unit{nM}       & improvement       & [5.2, 6.4]   & 0.74     & 2.02  & 1.69  \\
Incorp              & 13.1\;\unit{nM}      & 11.1\;\unit{nM}      & worsening         & [10.2, 12.0] & 0.62     & 3.33  & 2.80  \\
IncorpLowScav       & 11.7\;\unit{nM}      & 9.4\;\unit{nM}       & insignificant     & [8.6, 10.1]  & 0.61     & 2.35  & 2.08  \\
\bottomhline
\end{tabular}
\vskip 2mm
\caption{Statistics of \chem{[Si_{diss}]} (first two rows) and
        \chem{[Al_{diss}]} (other rows) of the simulations, in the West Atlantic
        GEOTRACES transect.  Here $D$ is the unweighted RMSD, $D_{\updownarrow}$
        the vertically homogenised RMSD (accompanied with its significance
        compared with $D_{\updownarrow}$ from the first row of each subtable,
        and the $2\sigma$ range), $r$ is the correlation coefficient, $RI$ the
        Reliability Index and $RI_{\updownarrow}$ is the vertically homogenised
        Reliability Index.  The significance is with respect to the first
        simulation in the corresponding subtable.}
\label{tab:statistics}
\end{table*}

\subsection{Reference simulation and observations}    \label{sec:alu_bg:base}
\begin{figure}
    \centering
    \includegraphics[width=\linewidth]{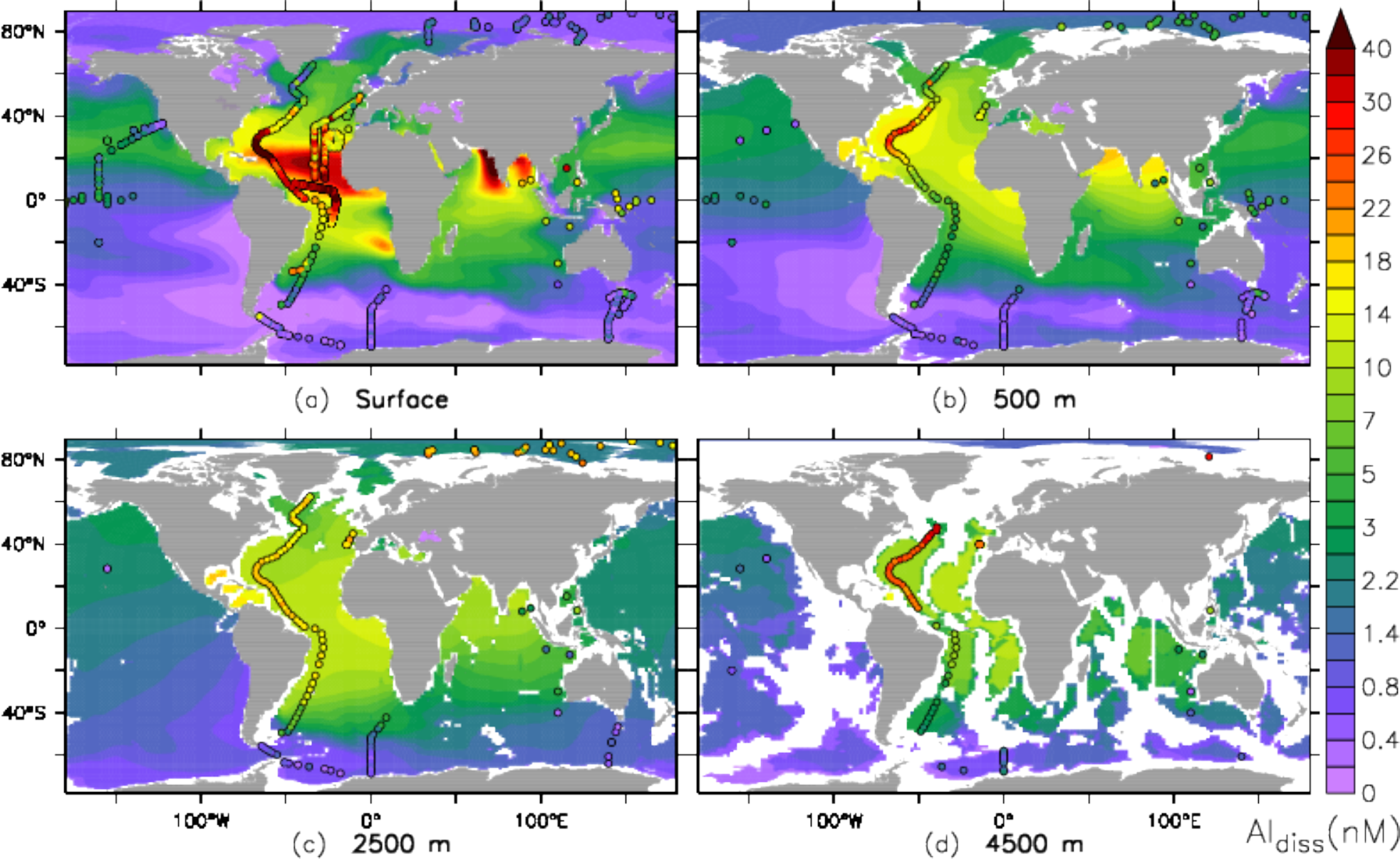}
    \caption{Average of the final model year (1750) of the dissolved aluminium
            concentration (nM) from the reference simulation (RefDyn2) at four
            depths.  The respective observations (same depth ranges) are
            presented as coloured dots.}
    \label{fig:ref_layers}
\end{figure}
\begin{figure}
    \centering
    \includegraphics[width=\linewidth]{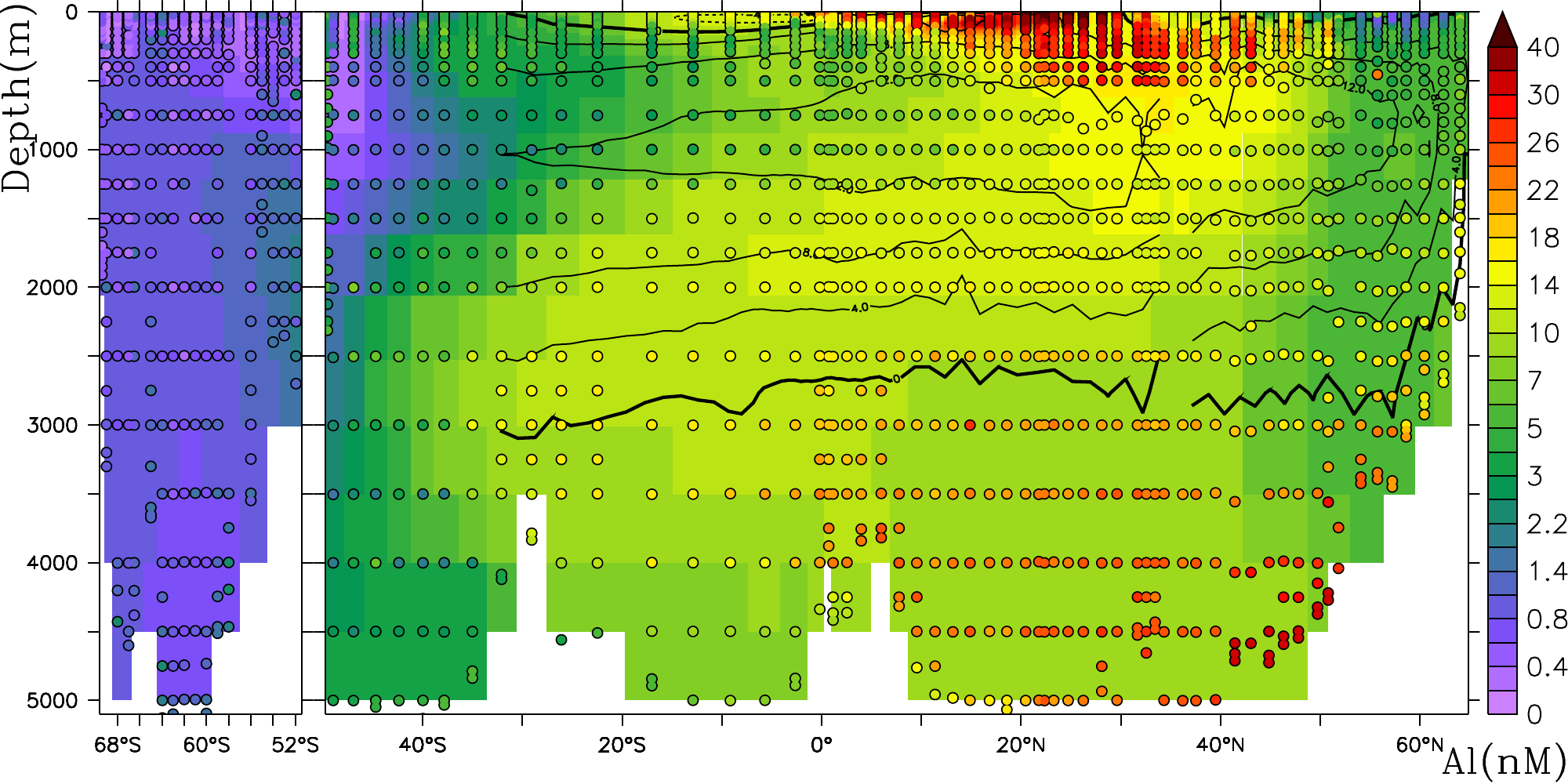}
    \caption{Dissolved aluminium (nM) from RefDyn2 (simulated year 1750) along
            the Zero Meridian and West Atlantic GEOTRACES transect.
            Observations are presented as coloured dots.  The contour is the
            Atlantic Overturning Stream Function (OSF), only defined north of
            Cape Agulhas and away from 36{\degree}\,N where cross-land mixing
            through the unresolved Strait of Gibraltar does not allow for a
            well-defined OSF.}
    \label{fig:ref_cruise}
\end{figure}
Figure~\ref{fig:ref_layers} shows the modelled yearly-average steady-state
\chem{Al_{diss}} concentration of RefDyn2 at four depths versus observations as
coloured dots.  In Fig.\ \ref{fig:ref_cruise} the GEOTRACES transects in the
West Atlantic Ocean and the Zero Meridian Southern Ocean are presented.  At the
surface of the Atlantic Ocean, the highest modelled and measured
\chem{[Al_{diss}]} is located from near the equator northwards to about
35{\degree}\,N.  In the polar oceans and in the South Pacific Ocean,
\chem{[Al_{diss}]} is very low.  At the West Atlantic transect, the MOC is
clearly reflected by the dissolved Al concentration.  The decrease of
\chem{[Al_{diss}]} from north to south in the North Atlantic Deep Water (NADW)
is due to net adsorptive scavenging onto settling biogenic silica.

The similarity between the model (RefDyn2) and the observations decreases in
the deeper North Atlantic Ocean, where according to the observations
\chem{[Al_{diss}]} increases with depth (for depths below 800\;m), while in the
model the dissolved Al concentration decreases with depth.  Besides this
general pattern of increasing \chem{[Al_{diss}]} with depth in the
observations, a very high concentration of \chem{Al_{diss}} is present between
45 and 50{\degree}\,N near the sediment, which enhances the dissimilarity
between the reference simulation and the observations.

\subsection{Improved dynamics}          \label{sec:alu_bg:dynamics}
In this reference simulation (RefDyn2) Dynamics~2 was used.  This forcing has
its maximum southward transport at a reasonably realistic depth of almost 3\;km
(contour in Fig.~\ref{fig:ref_cruise}), while \citet{vanhulten:alu_jms} used a
forcing with the strongest southward transport closer to 2\;km (Dynamics~1).
The more reasonable OSF depth results in an improved simulation of
\chem{[Al_{diss}]} in the West Atlantic Ocean (Fig.~\ref{fig:ref_cruise}).  The
RMSD of \chem{[Al_{diss}]} of the reference simulation from
\citet{vanhulten:alu_jms} (RefDyn1) versus observations is 8.3\,nM, while the
RMSD of this new simulation (RefDyn2) versus observations is 8.7\,nM.  The
difference between these RMSD values is insignificant
(Table~\ref{tab:statistics}).

The first two rows in Table~\ref{tab:statistics} present the goodness of fit
statistics for \chem{[Si_{diss}]}.  Inspecting the RMSD, RefDyn2 of this study
appears a significant improvement over RefDyn1 of \citet{vanhulten:alu_jms}.
The improved \chem{Si_{diss}} distribution gives credibility to the
\chem{[Si_{diss}]}-dependent sediment resuspension simulation (also:
Sect.~\ref{sec:alu_bg:discuss_underlying}).  Since \chem{[Si_{diss}]} is
improved, it is likely that the Si cycle as a whole is improved, but an
assessment of the Si cycle is beyond the scope of this paper.

\subsection{Sediment resuspension}      \label{sec:alu_bg:resuspension}
\begin{figure}
    \centering                          
    \includegraphics[width=.9\linewidth]{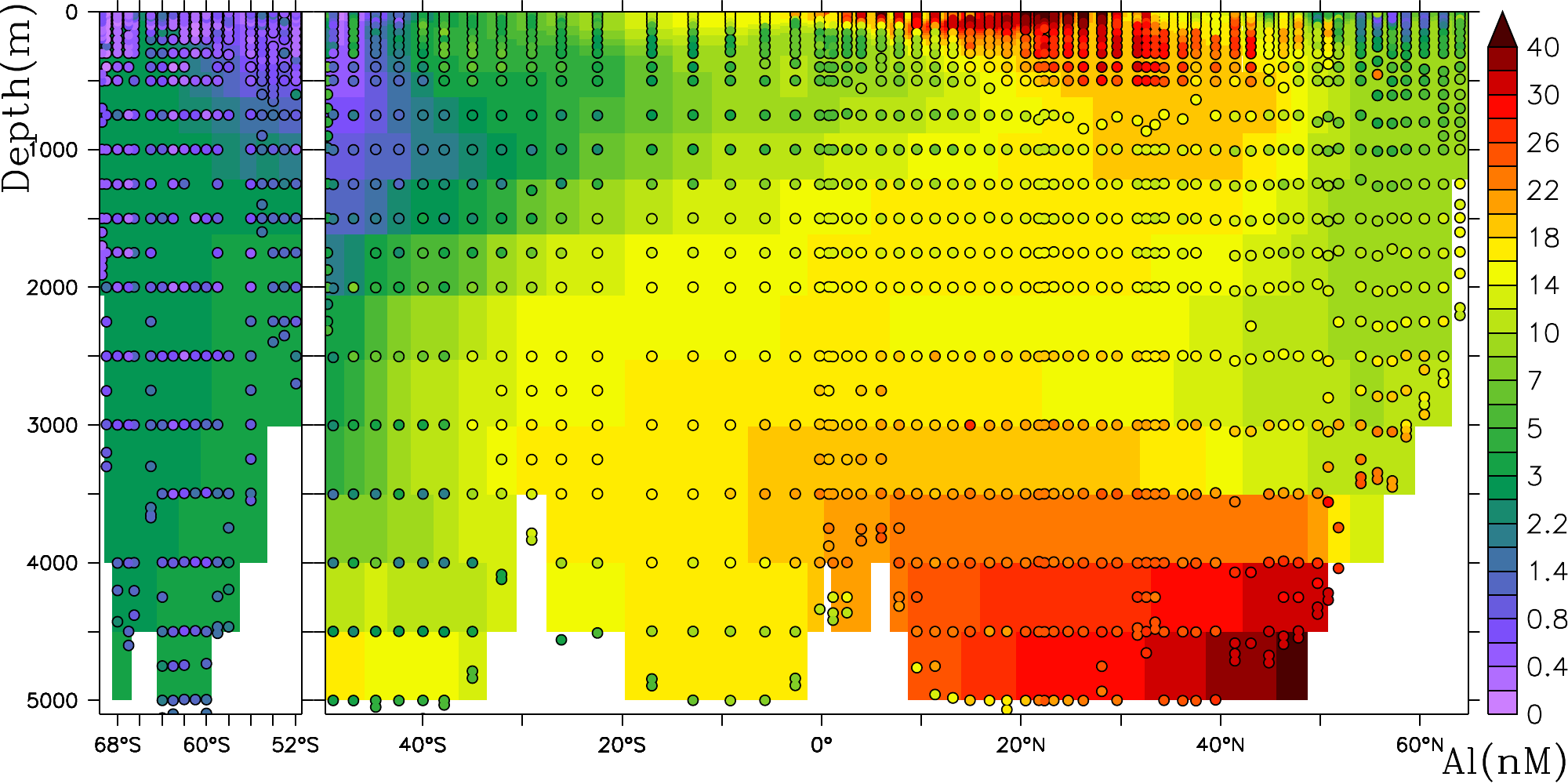}
    \caption{Simulated \chem{[Al_{diss}]}\;(\unit{nM}) from the
            \chem{[Si_{diss}]}-independent sediment
            resuspension simulation (500\;yr after forking) along the Zero
            Meridian and West Atlantic GEOTRACES transect (SedProp).
            Observations are represented by the coloured dots.}
    \label{fig:sed0_cruise}
\end{figure}

Figure \ref{fig:sed0_cruise} shows the \chem{[Al_{diss}]} resulting from the
simulation in which sediment resuspension is proportional to $w_s
\chem{Al_{ads}}/\Delta z_\text{bottom}$ just above the sediment (SedProp).
In the Northern Hemisphere, the deep \chem{[Al_{diss}]} is simulated much
better in this simulation than in RefDyn2 (Fig.~\ref{fig:ref_cruise}), but in
the Southern Hemisphere \chem{[Al_{diss}]} is simulated much worse.  The
dissolved Al concentration is too high near the bottom, and elevated
\chem{[Al_{diss}]} levels are found throughout the whole water column.
Nonetheless, the addition of resuspension in this way does significantly
improve the simulation ($D_{\updownarrow}$ = 5.2\;\unit{nM}, compared to
8.7\;\unit{nM} for the reference simulation).  However, in the southern
Atlantic Ocean and along the Zero Meridian in the Southern Ocean, sediment
input should be much closer to zero.

\begin{figure}
    \centering
    \begin{overpic}[width=.49\linewidth]{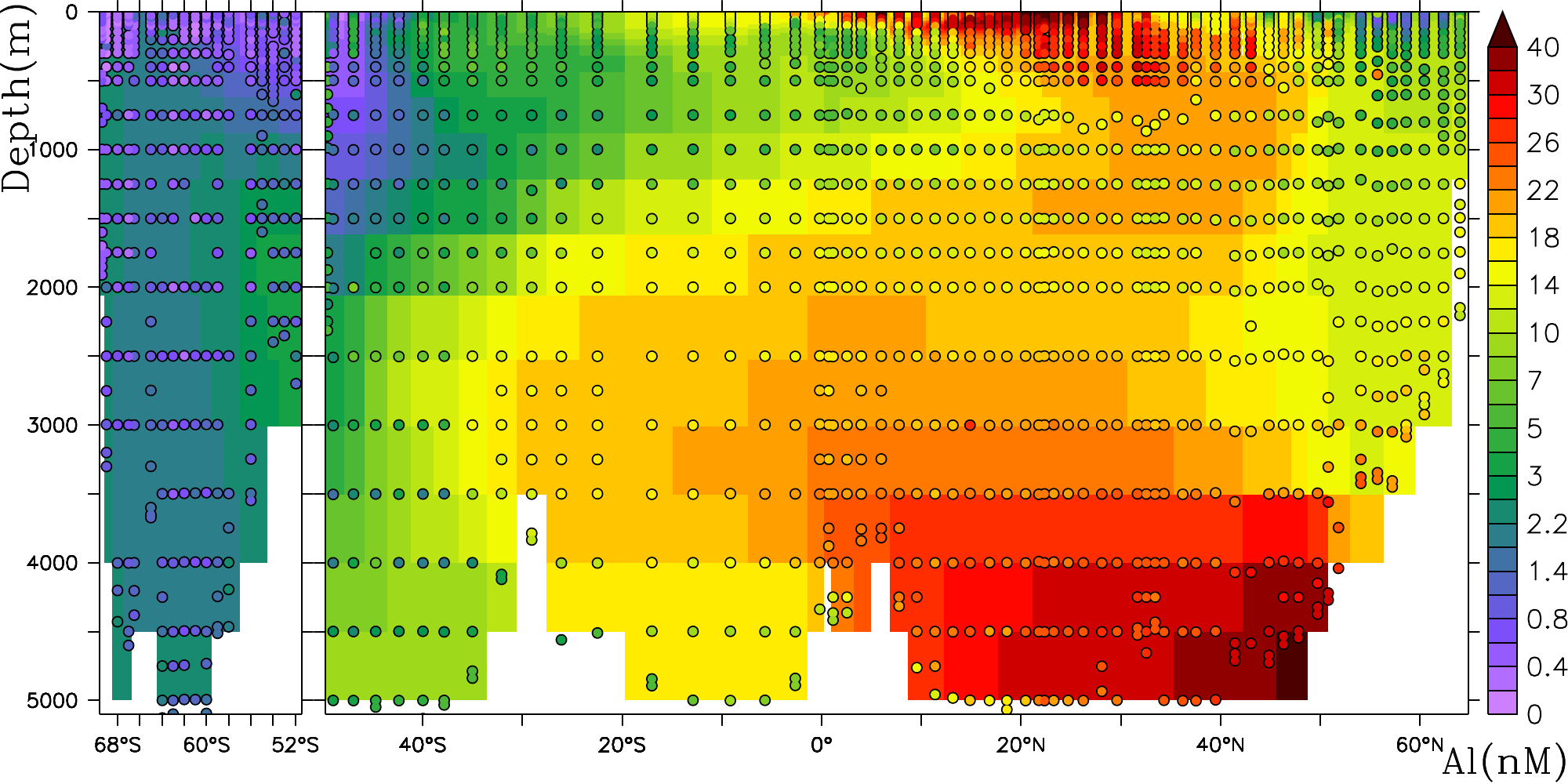}
        \put(87,8){\color{black}(a)}
    \end{overpic}
    \begin{overpic}[width=.49\linewidth]{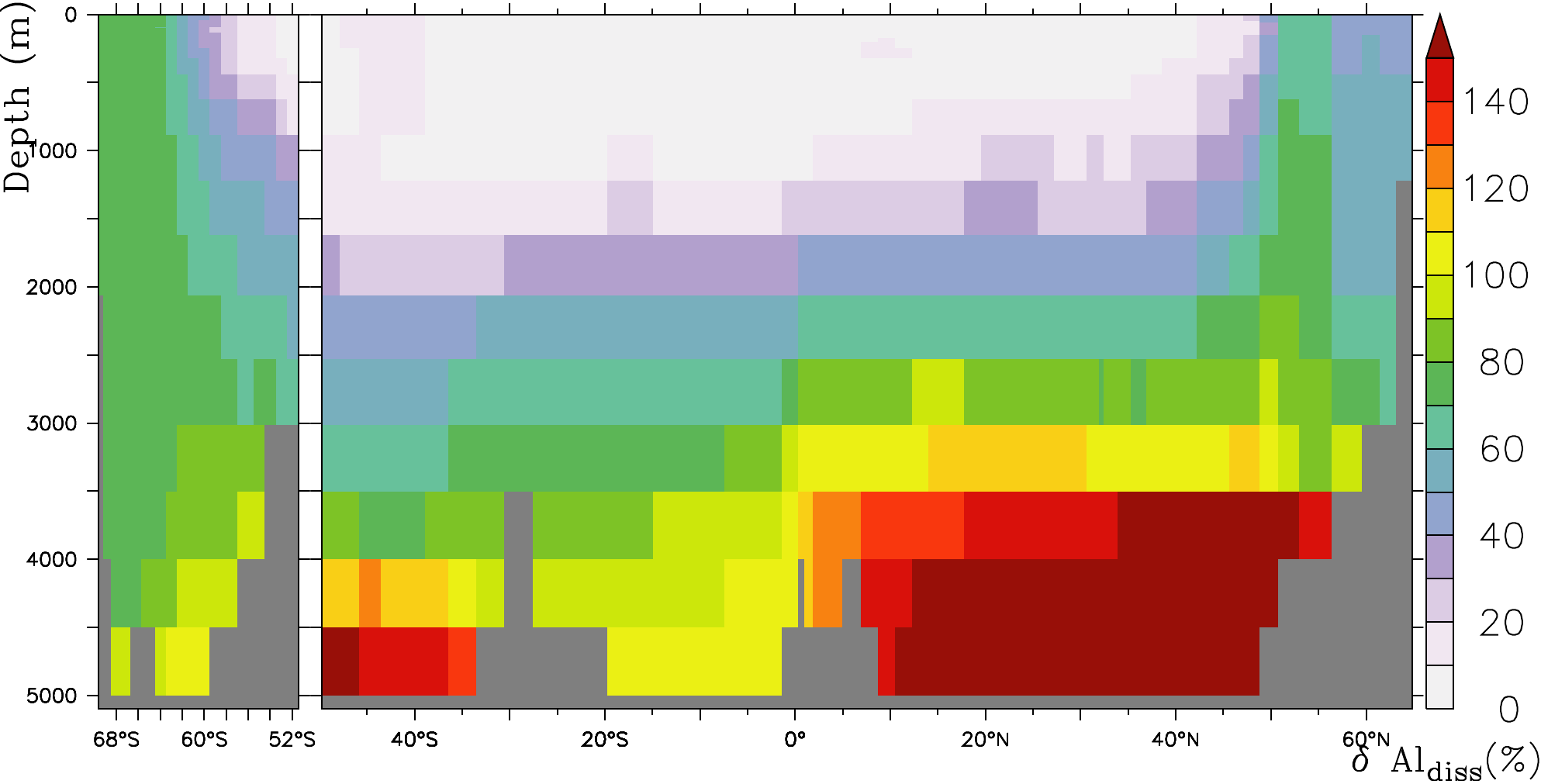}
        \put(87,8){\color{black}(b)}
    \end{overpic}
    \caption{Results from the \chem{[Si_{diss}]}-dependent sediment resuspension
            simulation following \citet{mackin1986} of year 500 after forking;
            sections along the Zero Meridian and West Atlantic GEOTRACES transects.
            (a) Modelled \chem{[Al_{diss}]}\,(nM) with observations plotted as coloured dots.
            (b) Relative difference of \chem{[Mn_{diss}]} between SedMackin and RefDyn2 ({\%}).}
    \label{fig:sed_cruise}
    \label{fig:sed_cruise_diff}
\end{figure}
Figure~\ref{fig:sed_cruise}a shows \chem{[Al_{diss}]} from simulation
SedMackin, where sedimentary Al addition depends on bottom water
\chem{[Si_{diss}]} according to \cite{mackin1986} ($\beta$ according to
Eq.~\eqref{eqn:beta_Mackin}).  As expected, also in SedMackin the sediment
resuspension source of Al results in a higher \chem{[Al_{diss}]} near
40--50{\degree}\,N in the deep North Atlantic Ocean
(Fig.~\ref{fig:sed_cruise_diff}b) compared to RefDyn2.  However, several
characteristics of the observations are better reproduced by SedMackin than
SedProp.  Firstly, in the southern hemisphere, the \chem{[Al_{diss}]} is only
slightly elevated near the sediment compared to the overlying water
(Fig.~\ref{fig:sed_cruise_diff}a).  The observations extend to practically the
bottom of the ocean, almost 6\,000\;m at some latitudes of the West Atlantic
GEOTRACES transect; while the model depth is only 5\,000\;m.  This makes a good
comparison of deep ocean \chem{[Al_{diss}]} between model and observations
difficult, but the slightly elevated \chem{[Al_{diss}]} in the near-sediment
observations is consistent with the slight elevation that is mainly confined to
the bottom model layer.  Secondly, the region from 0\degree\ to 45{\degree}\,N
at a depth below 2\;km has a higher \chem{[Al_{diss}]} compared to the
reference simulation (Fig.~\ref{fig:sed_cruise_diff}(b)) and better represents
the observations.

Based on the low \chem{[Si_{diss}]} alone, a high near-sediment
\chem{[Al_{diss}]} is expected north of 50{\degree}\,N as well.  Indeed, this
is the case in the observations, but it is not found in the model.  This could
be related to the relatively low resolution of the model which does not resolve
well the dynamical (advection and deep convection) and, related, biogeochemical
processes in this region.  As expected, the resuspended \chem{Al_{diss}} mixes
into the (lower) NADW, but most of it is scavenged again before reaching the
equator (Fig.~\ref{fig:sed_cruise_diff}(b)).  The overall resulting
\chem{[Al_{diss}]} is more consistent with the observations compared to the
original simulation without any sediment resuspension.  For the original
simulation, $D_{\updownarrow}=(8.8\pm 0.8)$\;\unit{nM}, while for the sediment
resuspension simulations, $D_{\updownarrow}=5.2$\;\unit{nM} and
$D_{\updownarrow}=5.9$\;\unit{nM}, which are statistically significant
improvements (see Table~\ref{tab:statistics}).

\begin{figure}
    \centering
    \includegraphics[width=.9\linewidth]{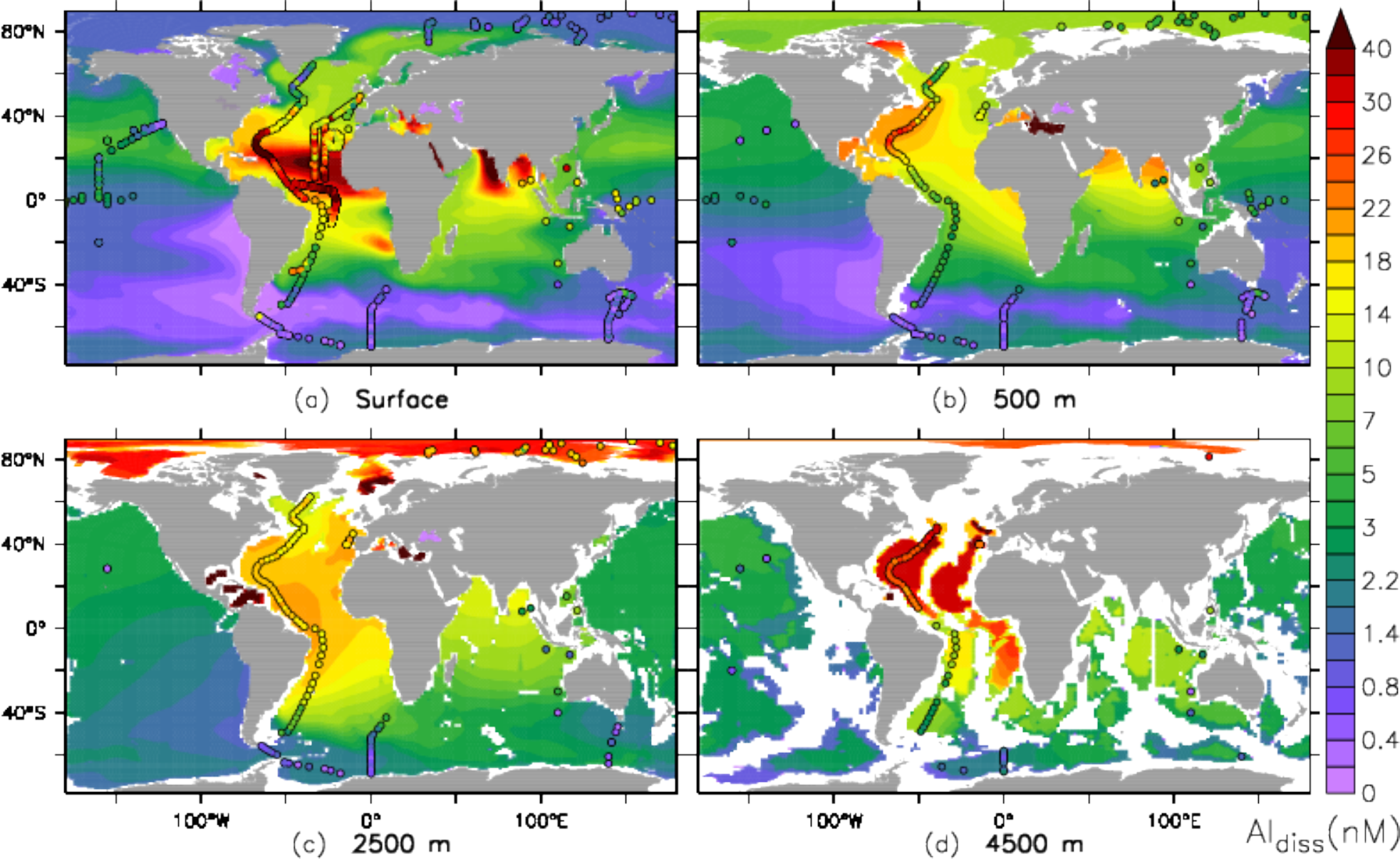}
    \caption{\chem{[Al_{diss}]}\;(\unit{nM}) of the \chem{[Si_{diss}]}-dependent
            simulation with sediment resuspension (SedMackin) at
            four depths (500\;yr after forking).}
    \label{fig:sed_layers}
\end{figure}
\begin{figure}
    \centering
    \includegraphics[width=.9\linewidth]{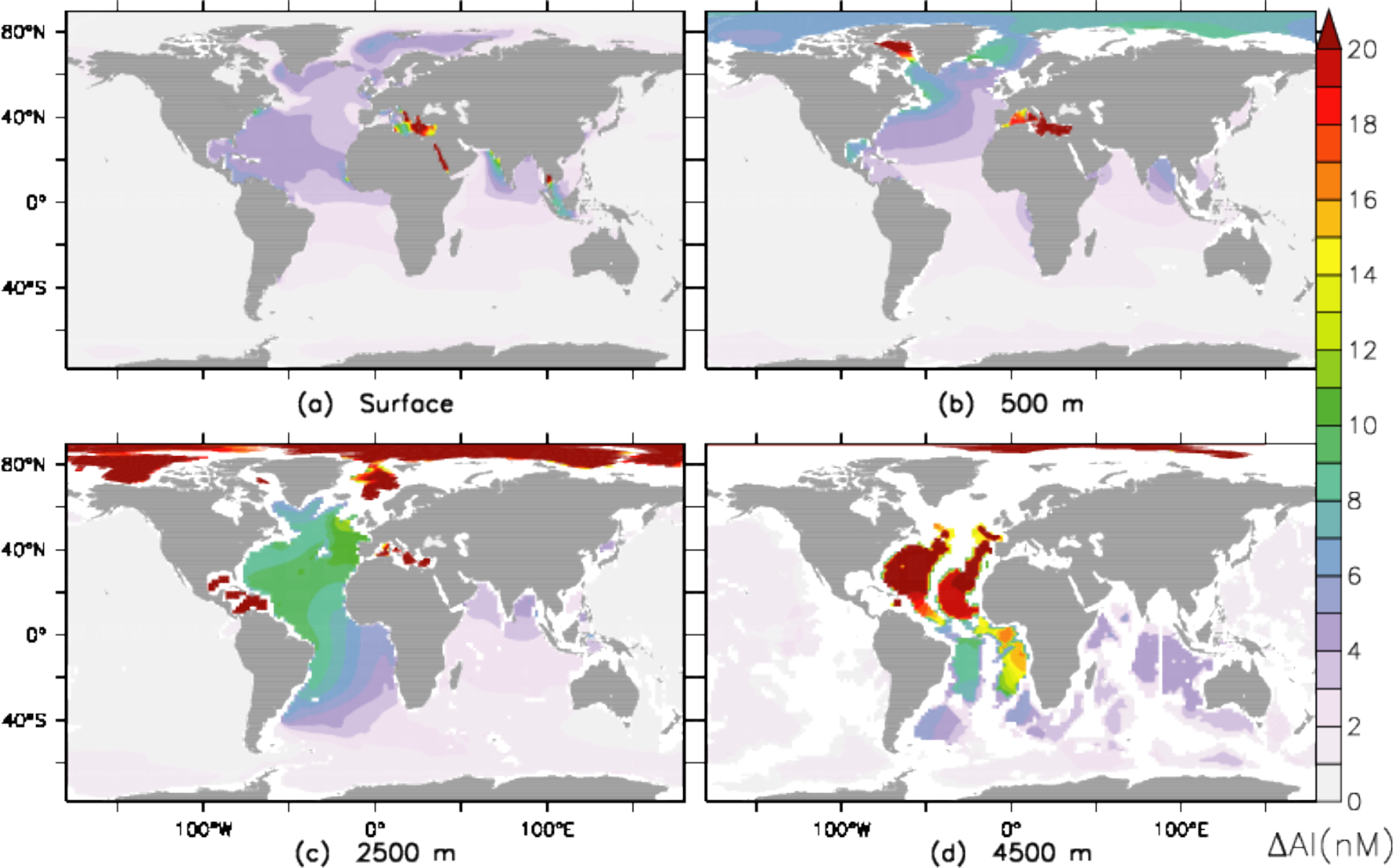}
    \caption{Difference of \chem{[Al_{diss}]}\;(\unit{nM}) between the
            simulation with \chem{[Si_{diss}]}-dependent sediment resuspension
            (SedMackin) and the reference simulation at four depths (500\;yr
            after forking).}
    \label{fig:sed_layers_diff}
\end{figure}
Figure~\ref{fig:sed_layers} shows the \chem{[Al_{diss}]} at four depths for
SedMackin, with observations as coloured dots.
Figure~\ref{fig:sed_layers_diff} shows the difference of \chem{[Al_{diss}]}
between this simulation and the reference simulation.  In several semi-enclosed
basins, like the Gulf of Mexico and the Arctic Ocean, and the Atlantic Ocean,
\chem{[Al_{diss}]} is higher compared to the reference simulation, especially
near the sediment.
%
%
The Gulf of Mexico, the Mediterranean Sea, Baffin Bay and the Arctic Ocean may
contribute to \chem{[Al_{diss}]} in the Atlantic Ocean.  However, the increase
of near-sediment \chem{[Al_{diss}]} in the West Atlantic Ocean at
45--50{\degree}\,N is much more likely caused by in situ resuspension and
subsequent dissolution (Fig.~\ref{fig:sed_cruise}(b)).

Several statistics show that SedMackin is an improvement over RefDyn2
(Table~\ref{tab:statistics}).  This, together with the presented concentration
plots, shows that the sediment redissolution process is an improvement of the
model.  This lends support to the hypothesis of a sediment source of Al in the
form of resuspension and subsequent dissolution in the real ocean.  In most of
the deep ocean (Figs~\ref{fig:sed_layers}c,d) the model overestimates the
observations.  However, the statistics do not show an improvement of SedMackin
over SedProp.  Apparently the choice of $\beta_0 = 16.85$ is too high, which
only showed up after a sufficiently long spin-up of the model.  It would
require several more trial and error model runs to arrive by iteration to the
optimal value of $\beta_0$, but this is beyond the scope of this paper.

\subsection{Biological incorporation}   \label{sec:alu_bg:incorporation}
\begin{figure}
    \centering
    \includegraphics[width=.9\linewidth]{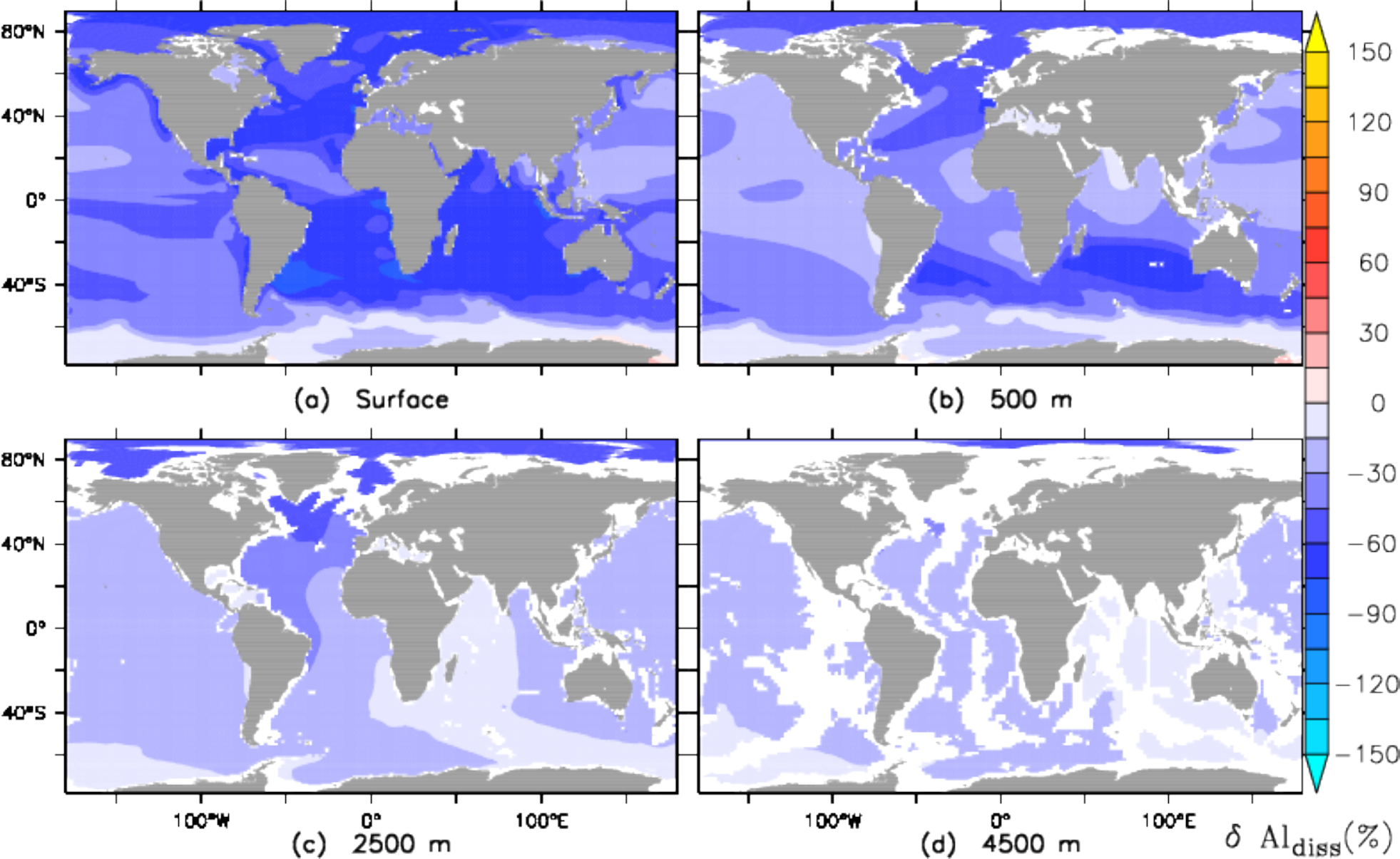}
    \caption{Relative difference of \chem{[Al_{diss}]} (\unit{\%}) between the
            simulation with biological incorporation (Incorp) and the reference
            simulation (RefDyn2) at four ocean depths (average over year 500).}
    \label{fig:inc_layers_reldiff}
\end{figure}

The relative difference between the simulated \chem{[Al_{diss}]} with and
without biological incorporation is presented in
Fig.~\ref{fig:inc_layers_reldiff} at four depths.  In the main thermocline,
north of 60{\degree}\,S, \chem{[Al_{diss}]} is significantly lower with
incorporation than without (Fig.~\ref{fig:inc_layers_reldiff} (a,b)).  While
the reference simulation RefDyn2 simulates the observed \chem{[Al_{diss}]} well
in the upper part of the ocean, in the simulation Incorp a large amount of Al
is removed by incorporation in addition to adsorptive scavenging from the upper
part of the ocean.

The significantly lower \chem{[Al_{diss}]} in the main thermocline in Incorp,
compared with RefDyn2, makes the modelled concentrations much lower than the
observed concentrations.  This suggests that incorporation may not occur (to
such an extent), which is consistent with the findings of \cite{vrieling1999}.
This result does not prove that biological incorporation of Al by diatoms does
not occur.  It only means that this way of incorporating Al into the frustules
in this model with its current configuration yields an unrealistically low
value of \chem{[Al_{diss}]}.  There are several possibilities to compensate
this effect so that future simulations may still be compatible with the
incorporation hypothesis.  Notably, the amount of Al incorporated into the
frustules may need to be significantly reduced, or the scavenging parameters
need to be adjusted, or dust dissolution must be increased.  The result of a
simulation with a decreased partition coefficient is presented next.  Further
discussion of the three options can be found in
Sect.~\ref{sec:alu_bg:discuss_incorp}.

\begin{figure}
    \centering                          
    \includegraphics[width=.9\linewidth]{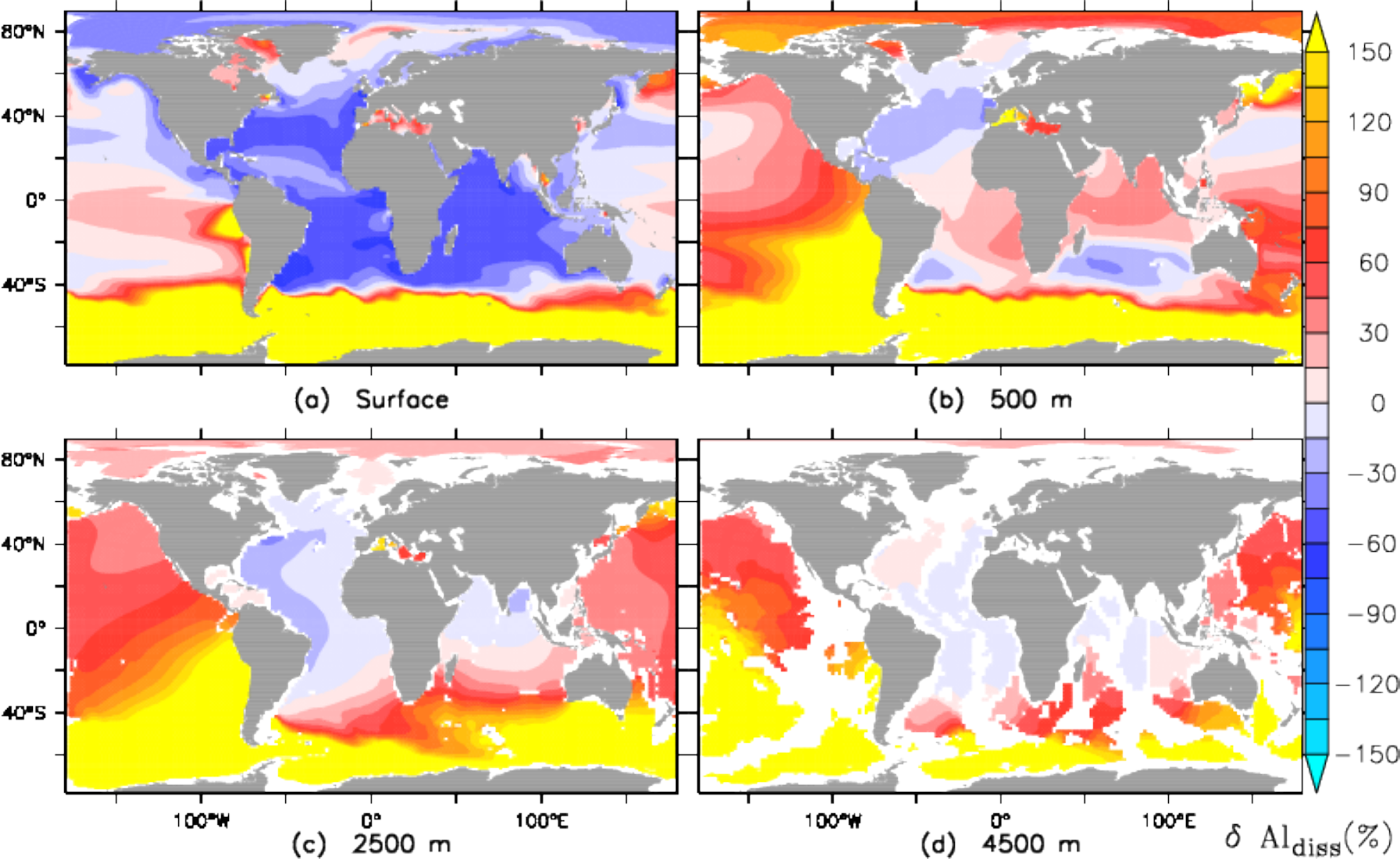}
    \caption{Result of a 500\;yr simulation with incorporation (no limitation,
            i.e.\ $R_\text{Al:Si} = \chem{[Al_{diss}]}/\chem{[Si_{diss}]}$) and
            a decreased scavenging partition coefficient $k_d' = k_d/4 = 28\cdot
            10^3$ dm$^3$\,mol$^{-1}$ (IncorpLowScav).  Relative difference of
            \chem{[Al_{diss}]} between IncorpLowScav and RefDyn2.}
    \label{fig:incorp1kd_layers_reldiff}
\end{figure}

Figure~\ref{fig:incorp1kd_layers_reldiff} presents the relative concentration
difference between IncorpLowScav and RefDyn2.  The decrease in the Atlantic
Ocean surface waters is slightly smaller than in the case with incorporation
and a high $k_d$ (compare with Fig.~\ref{fig:inc_layers_reldiff}).  However, in
the Southern Ocean the \chem{[Al_{diss}]} has increased relatively by a
considerable amount, yielding concentrations much higher than the observed
concentrations.

The RMSD between IncorpLowScav and the observations (at the West Atlantic
GEOTRACES transect) is 9.4\;\unit{nM}.  Even though this is a significant
improvement compared to Incorp, it is only the case when absolute residuals are
considered (RMSD).  When the correlation coefficient is considered,
IncorpLowScav ($r=0.61$) appears not to be an improvement to Incorp ($r=0.62$).
In fact, the correlation coefficient for IncorpLowScav is the lowest of all
simulations.  Indeed, IncorpLowScav shows poor model performance in the
Southern Hemisphere (Fig.~\ref{fig:inc_layers_reldiff}).  Compared to the
simulation without incorporation (RefDyn2), the run with a incorporation with
fourfold lower partition coefficient coefficient (IncorpLowScav) performs
poorly.  This degradation in model performance is insignificant when the RMSD
is considered (which considers absolute residuals).  Dimensionless goodness of
fits like the Reliability Index $R\!I$ or the correlation coefficient $r$ do
not show any significant improvement or worsening of IncorpLowScav either,
compared with RefDyn2 (significance values not presented).  Finally, decreasing
the first-order rate constant $\kappa$ has a very similar effect as decreasing
$k_d$ (results not presented).

From these considerations it may be concluded, within the limitations of the
model, that if incorporation is an important process at all, it is unlikely to
occur proportional to the ambient \chem{[Al_{diss}]}/\chem{[Si_{diss}]} ratio
in surface seawater but rather in a much smaller ratio (i.e.\ $c_\text{in}\ll
1$ in Eq.~\eqref{eqn:xi} as further discussed in
Sect.~\ref{sec:alu_bg:discuss_biological}).

\section{Discussion}                    \label{sec:alu_bg:discuss}
\subsection{General biogeochemistry}
\subsubsection{Underlying model}        \label{sec:alu_bg:discuss_underlying}
\begin{figure}[ht!]
    \centering
    \subfloat[Dynamics~1]{
    \includegraphics[width=.5\linewidth]{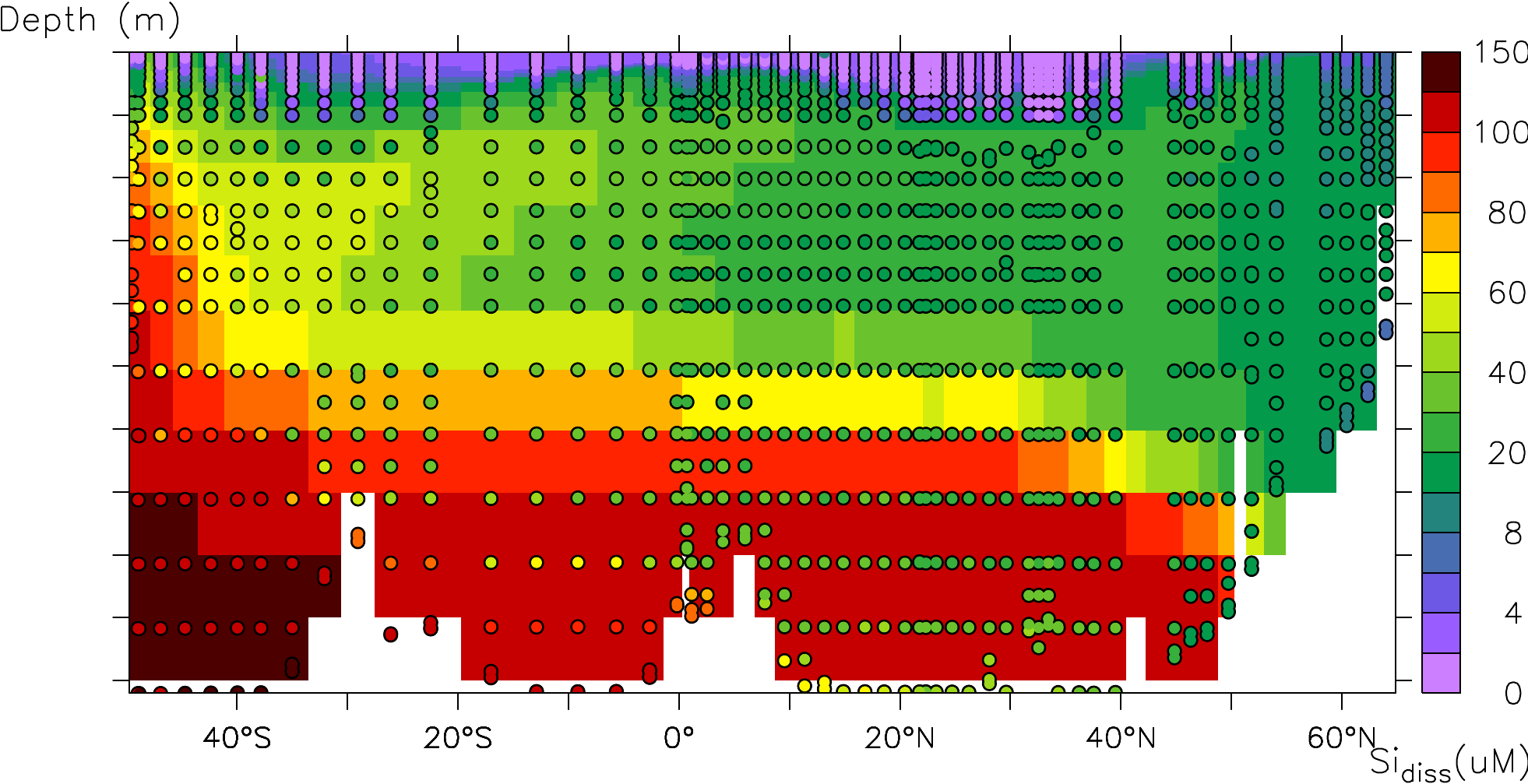}
        \label{fig:Si_dyn1}
    }
    \subfloat[Dynamics~2]{
        \includegraphics[width=.5\linewidth]{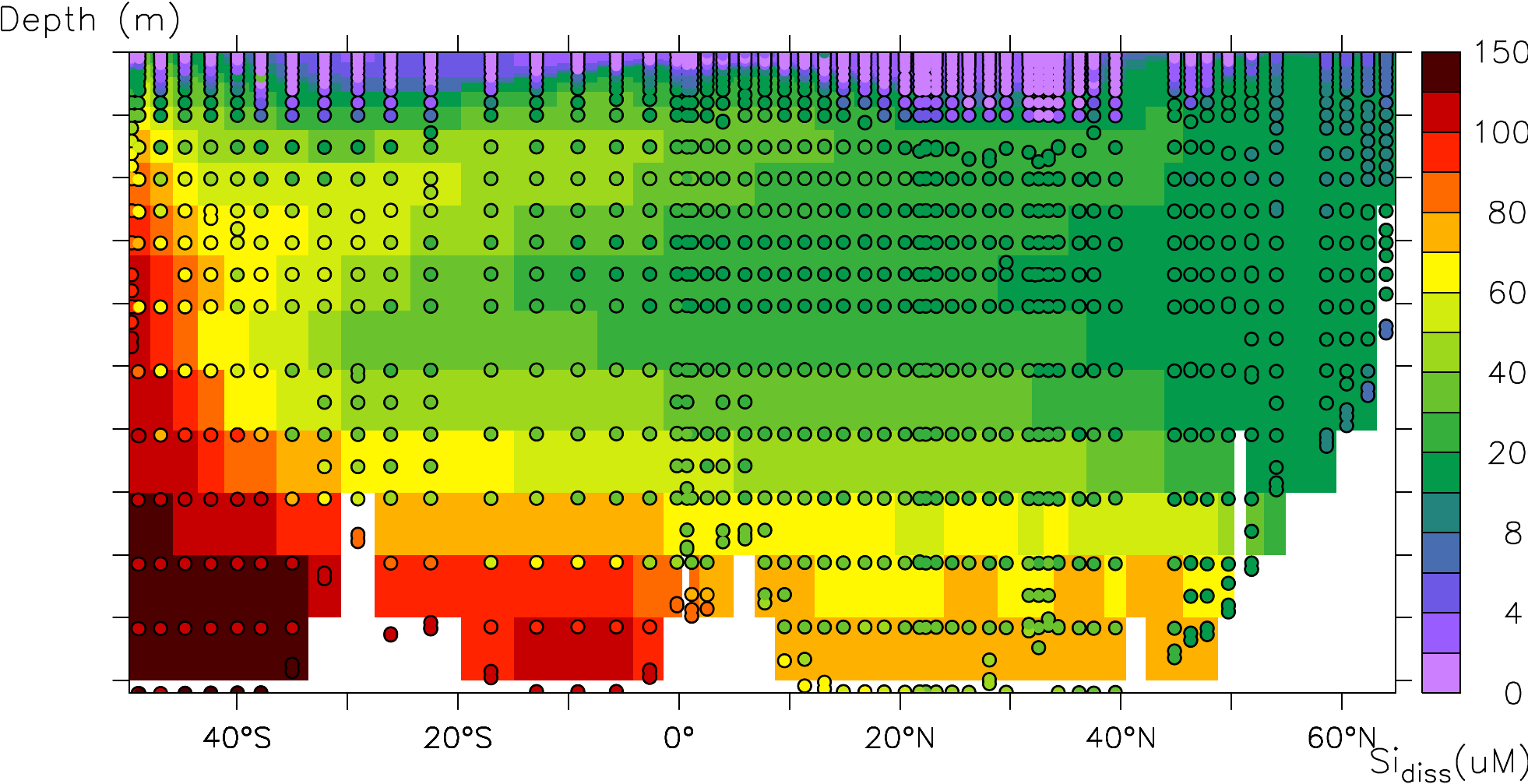}
        \label{fig:Si_dyn2}
    }
    \caption{The modelled silicic acid concentration at the West Atlantic
            GEOTRACES section.  Coloured dots are measurements (data from
            K.~Bakker, E.~van~Weerlee, M.~Rijkenberg and H.J.W.~de~Baar).  Both
            simulated and observed concentrations are in \unit{\mu M} (on a
            non-linear scale).}
    \label{fig:Si_sections}
\end{figure}

Clearly, besides the dust solubility, the scavenging parameters and
resuspension parameterisation, the Al model depends on the dynamics
(Sect.~\ref{sec:alu_bg:dynamics}) and the underlying biogeochemical model as
well.  Both the dynamics and the biogeochemical model have a strong impact on
[\chem{Si_{diss}}].  Figure~\ref{fig:Si_sections} presents modelled
[\chem{Si_{diss}}] for both dynamical fields, with measured [\chem{Si_{diss}}]
as coloured dots.
Clearly Fig.~\ref{fig:Si_dyn1}, corresponding with Dynamics~1, shows a strongly
overestimated [\chem{Si_{diss}}] in the deep ocean, while this overestimation
is significantly reduced in Fig.~\ref{fig:Si_dyn2} (Dynamics~2).  The reason
for this is that in Dynamics~2 the northward flux of AABW is smaller and does
not go as far northwards as Dynamics~1.  Hence, less \chem{Si_{diss}}, which is
rich in AABW, reaches the northern parts of the deep Atlantic Ocean.  In other
words, the more realistic deep overturning cell in Dynamics~2 results in a more
realistic simulation of [\chem{Si_{diss}}] in the deep Atlantic Ocean (see
Table~\ref{tab:statistics}).  Even though we have improved the \chem{Si_{diss}}
distribution by using a different dynamical forcing, the absolute value of the
deep [\chem{Si_{diss}}] does not match very well the observations.  At most
places in the deep West Atlantic Ocean, \chem{[Si_{diss}]} still overestimates
the observations (Fig.~\ref{fig:Si_sections}(b)).  For the sediment
resuspension simulation SedMackin, this has been taken into account by
modifying the proportionality factor $\beta_0$ (see Eq.~\eqref{eqn:beta_Mackin}).
Still, $\beta_0$ was estimated somewhat too high as noted above.

The \chem{Si_{biog}} concentration is important as well for the Al model,
namely for scavenging.  However, there is no consistent observational dataset
of [\chem{Si_{biog}}].  Available data \citep[e.g.][]{lam2011} generally report
particulate silica as the sum of biogenic silica (diatom frustules) and silica
mineral phases such as silt and clay particles.  For this reason, no one-to-one
comparison between modelled and measured biogenic silica can be performed; only
a qualitative analysis can easily be done.  \cite{aumont2006} show that primary
production of diatoms is small in the oligotrophic North Atlantic Ocean,
consistent with measurement-derived biogenic silica export \citep[e.g.][Color
Plate 4]{sarmiento2006}.

\subsubsection{Dust dissolution}
Fractions of each of the elements in lithogenic dust particles dissolve in the
upper mixed layer, while below the mixed layer the dissolution of the elements
from dust is smaller.  It is generally assumed that most of the dust is
refractory.  Therefore, in our model, Al from lithogenic particles is not
dissolved below the mixed layer.  The neglect of dissolution below the mixed
layer is often casually assumed and has been suggested by many studies on Al in
the ocean \citep[e.g.][]{orians1986,baker2006,buck2006}.  However, no explicit
observational study has been published on the fate of lithogenic dust particles
in the water column below the mixed layer.

One of the sensitivity simulations of \citet{vanhulten:alu_jms} suggests that
there is no dissolution below the mixed layer.  That simulation, DustSS,
included dissolution of dust below the surface layer, while there was no
sediment source.  DustSS did not perform better compared to their reference
experiment, RefDyn1.  The impact of the dissolution below the mixed layer had
especially an aggravated impact in the West Atlantic Ocean between 500 and
1500\,m depth.  Hence, there is good reason for excluding water column
dissolution of lithogenic Al.  The sediment resuspension simulations of this
study (SedProp and SedMackin) include a deep source of \chem{Al_{diss}}.  If
dissolution below the mixed layer would be added to these simulations, this
would probably worsen the \chem{Al_{diss}} distribution.  Hence, there is good
reason for excluding water column dissolution of lithogenic Al.

\subsubsection{Reversible scavenging}
In our model \chem{Al_{diss}} is reversibly scavenged.  This means that \chem{Al_{diss}}
is scavenged and, during downward settling, partly released in the water
column.  The release is caused by a combination of decreasing \chem{Si_{biog}}
and \chem{Al_{diss}} (Eqs~\eqref{eqn:scav} and~\ref{eqn:equil}).  At first sight,
this release appears necessary to explain the non-zero water column
concentrations of \chem{Al_{diss}}.  This is consistent with the notion that
\chem{Si_{biog}} with \chem{Al_{ads}} follows the MOC and slowly releases
\chem{Al_{diss}}.  This can be seen in Fig.~\ref{fig:observations}(a) and
Figs.~\ref{fig:ref_layers} and \ref{fig:sed_layers}, where \chem{[Al_{diss}]}
shows the `imprint' of the MOC.  Still, it is possible that desorption occurs
at a slower rate than adsorption, and that desorption hardly depends on the
particle concentration \citep{moran1992:kinetics}.

In the real ocean adsorptive scavengers other than biogenic Si, as well as
different Al particles, may play a role.  For example, colloidal Al and other
colloids facilitate the removal of trace metals like Al
\citep{moran1989,moran1992:colloids}.  However, the fraction of colloidal Al is
only 0.2--3.4\,{\%} (of the operationally defined filtrate $< 0.2$\;\unit{\mu
m} \chem{Al_{diss}} pool), meaning that the removal of Al occurs mainly through
direct adsorption, or the colloidal fraction is rapidly turned over
\citep{dammshauser2012}.  Either case implies that the model may skip the
colloidal fraction and convert \chem{Al_{diss}} to \chem{Al_{ads}} in a single
step.\footnote{Furthermore, \cite{moran1988:evidence} note that Al has a large truly
dissolved fraction, and ``will not be removed as effectively [as Th] during the
coagulation process.'' Finally, and also because of these considerations, the dataset
available now does not distinguish the colloidal from the soluble pool.}


\subsection{Sediment source}     \label{sec:alu_bg:discuss_resusp}
\subsubsection{Resuspension versus diffusion}
Typically in shallow waters, dissolved iron (Fe) diffuses from the sediment
into the water above due to build-up of reduced Fe within sediment pore waters.
\cite{vanhulten:alu_jms} performed a simulation with a sediment source of
dissolved Al analogous to Fe.  However, reduction and subsequent diffusion of
Al from sediments does not typically occur.  The sediment input simulation in
this work is based on sediment resuspension in deep ocean bottom waters instead
of upward diffusion across the sediment-water interface.

The resuspension hypothesis has a stronger observational basis than the
diffusion hypothesis.  Observations show that there appears to be an Al source
from sediments where there is resuspension, e.g.\ around the Grand Banks
\citep{moran1991}.  This typically coincides with regions of deep water
formation or it occurs around seafloor elevations.  When there does not appear
to be resuspension, there is generally no source of \chem{Al_{diss}}.  This of
course does not exclude the possibility of diffusion of Al out of sediments,
but the observations link near-sediment Al elevations to regions of
resuspension.  Further discussion and references may be found in
\cite{middag:prep}.

\subsubsection{Shortcomings}    \label{sec:alu_bg:discuss_resusp_short}
Sediment resuspension is actually induced by near-sediment turbulence, creating
a~layer of water above the sediment containing significant amounts of suspended
sediment particles.  This layer is about 200 to 1000\,m thick and is referred
to as the \emph{nepheloid layer} \citep{jackson2005}.  Due to the low vertical
resolution of the model (bottom layer is up to 500\;\unit{m} thick), it was not
possible to parameterise sediment resuspension by means of the turbulence
parameters present in the model.  Instead, a~certain portion of sedimented
biogenic silica was redissolved into the bottom model layer.  This is hence not
a mechanistic model but a simple parameterisation.  This is possible because
the bottom layer thickness is of the same order of magnitude as the nepheloid
layer.

There are no observations or good estimates of Al flux from sediment
resuspension, and there is no sediment model for Al available.  Therefore it is
assumed that recently sedimented \chem{Al_{ads}} is resuspended and
subsequently partly dissolved \citep[e.g.][]{lampitt1985,hwang2010}.  Besides
\chem{Al_{ads}} there is another particulate Al tracer present in some of the
model simulations, namely \chem{Al_{biog}}.  This is not used here, since Al
incorporated into the frustule lattice is harder to dissolve than Al adsorbed
onto \chem{Si_{biog}}.  Also, \citet{koning2007} state that a significant
amount of the Al, associated to diatom frustules post-mortem, can be rinsed of
and is thus bound to the surface.

\subsubsection{Inhibition}
The dissolution of Al from resuspended sediments appears to be inhibited by
\chem{Si_{diss}}.  This effect has been included in Eq.~\eqref{eqn:beta_Mackin}.
Here we try to give an extensive qualitative rationale for the hypothesis,
while a derivation of the final equation can be found in
Appendix~\ref{sec:alu_bg:derivation}\@.  We believe the elevation of
\chem{[Al_{diss}]} in bottom waters (Fig.~\ref{fig:observations}(a)) is due to
significant dissolution/desorption from resuspended sediment in part of the
northern hemisphere, a process apparently inhibited in the Southern Hemisphere.
This may be explained by the mirror image distribution of dissolved silicate
(Fig.~\ref{fig:observations}(b)).  Namely, the \chem{Si_{diss}} concentration is very high in the AABW in the southern
hemisphere and flows as far north as $\sim$40{\degree}\,N, albeit somewhat
diluted by vertical mixing with overlying NADW that has lower \chem{Si_{diss}}
concentrations.  The high concentration of \chem{Si_{diss}} in bottom waters
possibly prevents the dissolution/desorption of Al from sediments. Briefly, in
the northernmost part of the transect the DSOW cascades down over the seafloor
from $\sim\,65${\degree}\,N to $\sim\,45${\degree}\,N, underway acquiring more
and more dissolved Al due to redissolution/desorption from resuspended sediment
particles in the bottom nepheloid layer, which is very thick here
\citep{biscaye1977,gross1988}. Indeed, from 65\degree N to 45{\degree}\,N the
dissolved Al in the deepest bottom water sample increases from 12\;\unit{nM} at
65{\degree}\,N to 34 nM at 44.8{\degree}\,N \citep[][their Suppl.\
Figure~S-3]{middag:prep}.  The parallel increase of \chem{[Si_{diss}]} is
slightly more modest from 8\;\unit{\mu M} at 65{\degree}\,N to 21\;\unit{\mu M}
at 44.8{\degree}\,N. However, somewhere in between 45{\degree}\,N and
40{\degree}\,N the southward flowing DSOW becomes underlain by the northward
extreme of AABW with much higher $\chem{[Si_{diss}]} \approx 45$\;\unit{\mu M}
at 39.5{\degree}\,N.  There is still significant sediment resuspension based on
optical back scatter observations, but no more release of Al, hence
\chem{Si_{diss}} appears to inhibit any further dissolution/desorption of Al
from resuspended sediments.

The overall sediment source of \chem{Al_{diss}} is likely due to a combination
of processes (bottom current velocity, resuspension, partial dissolution of
clay minerals, partial dissolution of biogenic debris (\chem{Si_{biog}},
\chem{Al_{biog}}) and desorption) where porewater chemistry within the
sediments likely plays a role as well.  However, neither the observations
(Fig.~\ref{fig:observations}a) nor the simulation modelling of dissolved Al
have the adequate vertical resolution to resolve all these processes. Instead
we follow \cite{mackin1986} and the derivation in Appendix
\ref{sec:alu_bg:derivation}.

\subsection{Incorporation}              \label{sec:alu_bg:discuss_incorp}
\subsubsection{Biological}              \label{sec:alu_bg:discuss_biological}
The simulation Incorp resulted in a much too small [\chem{Al_{diss}}] in the
upper 500\;m of the ocean, especially below (and downstream of) dust
deposition.  In this section three possibilities are discussed that may leave
open the option of biological incorporation.

Firstly, too much Al might be biologically incorporated into the frustules.  A
different moderation term $R_\text{Al:Si}$ would be able to change that, and
one way is to decrease $c_\text{in}$.  This parameter describes a general
preference of silicic acid above aluminium when it is smaller than one and vice
versa.  \cite{han2008} defined $c_\text{in} = 0.08845$, loosely based on
\cite{gehlen2002}.  Alternatively, one may also define a maximum incorporation
$r_\text{max}$.  This parameter signifies that diatoms do not allow aluminium
in their frustule above a certain percentage.  Several equivocal values for
$r_\text{max}$ may be derived from different studies, among which $0.007$,
based on \cite{gehlen2002} as well, or 0.01 \citep{vancappellen2002}, or
0.0022, an incorporation ratio based on observations of \chem{[Al_{diss}]} and
\chem{[Si_{diss}]} at remineralisation depth \citep{middag2009}.  An additional
simulation was performed with the parameterisation by \cite{han2008}, i.e.\
with $c_\text{in} = 0.08845$ and $r_\text{max}=\infty$ (results not presented
here).  This simulation does not yield significant changes compared to the
reference simulation (RefDyn2).  Hence, from these simulations it cannot be
concluded whether incorporation (with a reduced incorporation rate
$c_\text{in}$) occurs.

Secondly, whether incorporation is moderated or not, the part of the model that
is independent of incorporation is likely to need retuning, since incorporation
functions as an extra sink of Al.  The settling velocity $w_s$ and the other
scavenging parameters $k_d$ and $\kappa$ may need to be adjusted (see
Eqs~\eqref{eqn:scav}, \ref{eqn:equil} and \ref{eqn:sink}).  For instance, the
partition coefficient $k_d$ may be decreased.  This leaves more
\chem{Al_{diss}} in the surface ocean, compensating the decrease of
\chem{[Al_{diss}]} by incorporation.  To test this hypothesis a simulation was
performed with incorporation and a decreased $k_d$ (IncorpLowScav).  This
simulation showed a large effect of $k_d$ on the Southern Ocean, and not on the
Atlantic Ocean, where incorporation had such a large effect
(Sect.~\ref{sec:alu_bg:incorporation}).  Why is this the case?

In our model, in chemical equilibrium, the adsorbed Al concentration is
proportional to both \chem{[Al_{diss}]} and \chem{[Si_{biog}]}
(Eq.~\eqref{eqn:equil}).  The effect of the biogenic Si dependence is evident
in high-\chem{[Si_{biog}]} areas like the Southern Ocean.  Furthermore,
\chem{Si_{biog}} is present in and below the photic zone, especially in the
Southern Ocean.  This results in scavenging throughout a significant portion of
the water column.  In other words, $k_d$ has a much stronger sensitivity in the
polar oceans compared to other locations, as is actually shown by a simulation
of \citet{vanhulten:alu_jms}.  A smaller $k_d$ therefore results in more
\chem{Al_{diss}} in the Southern Ocean, while it has little effect on the more
problematic incorporation-induced decrease of \chem{[Al_{diss}]} in the
Atlantic Ocean.

Increasing dust deposition, or solubility, is another possibility to compensate
the decrease of \chem{[Al_{diss}]} in the main thermocline in the simulations
where biological incorporation occurs.  A preliminary simulation suggests that
more dust only affects the very surface of the ocean.  Dust deposition and
dissolution of Al in the surface layer does not affect the subsurface
\chem{[Al_{diss}]} much (at 50--400\;m depth), while as a consequence of
biological incorporation a strong depletion of \chem{Al_{diss}} occurs there as
well (result not presented here).  Dissolving lithogenic dust particles below
the ocean surface is another option.  This has been done in
\citet{vanhulten:alu_jms} by means of instantaneous dissolution in the water
column upon deposition on the sea surface according to a function that
exponentially decreases with depth.  The depth-dependent dissolution function
can be fitted to the observations or the reference simulation, giving a
reasonable first-order simulation of \chem{[Al_{diss}]}.  Obviously, a
simulation with incorporation and a dissolution function fitted to the
observations is not very strong evidence for incorporation.

To summarise, the simulation IncorpLowScav showed that changing $k_d$ cannot
solve the problem, and we have to understand why this is the case.  Also a
simulation with increased dust dissolution does not give much hope.  This
suggests that biological incorporation of Al into diatoms has only a
second-order effect on the dissolved Al distribution. This is in line with the
likely overestimation of Al to Si ratio in our model during incorporation as
discussed near the beginning of this section.  Indeed, there are several
studies that cannot detect the process of biological incorporation of Al into
living diatoms \citep[e.g.][]{vrieling1999} and many studies are unclear about
the actual mechanism of removal \citep[e.g.][]{moran1988:evidence,ren2011}.
However, as discussed, at the same time there are many studies strongly
suggesting the biological incorporation.  Either parts of the model (which is
of relatively large complexity and contains many degrees of freedom) are
overlooked, e.g.\ in the complexities of the Si cycle; or the amount of
incorporation is smaller than assumed in Incorp.

\subsubsection{Post-mortem diagenesis within the sediments}
\cite{koning2007} and \cite{loucaides2012,loucaides2010} have suggested that
most of the Al found in diatom silicate in sediments is incorporated after
burial. These papers indicate that the Al/Si ratio in living diatoms is most
likely considerably lower than the estimates used in the incorporation
simulations.  Most incorporation is indeed after burial and hence post-mortem.
However, sedimentary processes are not the focus of this work and only matters
are discussed that are of direct importance for the processes in the water
column.  What we aimed for here is to test the effect on [\chem{Al_{diss}}] by
including biological incorporation in opal of growing diatoms by using the
upper limit of Al/Si (namely the ambient dissolved Al/Si concentration ratio in
the surface ocean waters). A study on sedimentary processes is beyond the scope
of this paper.

\conclusions    \label{sec:alu_bg:conclusion}
The \chem{Al_{diss}} distribution in the upper part of the ocean has previously
been simulated reasonably well with only a dust source and reversible
scavenging as the removal process \cite{vanhulten:alu_jms}.  However, the
\chem{[Al_{diss}]} was strongly underestimated in the deep North Atlantic
Ocean, highlighting deficiencies in this model.  The simulation is
significantly improved by the use of different dynamical fields and the
addition of a resuspension source.  The latter supports the idea that the most
significant sources of Al to the ocean are dust deposition and sediment
resuspension, and the most important internal process is likely to be
adsorptive scavenging.  The Al release from resuspended sediment appears to
depend on both \chem{Al_{ads}} sedimentation and bottom water
\chem{[Si_{diss}]}.  It has been shown that a parameterisation based on
\cite{mackin1986} is able to simulate the deep ocean \chem{[Al_{diss}]}
realistically, supporting the idea of stoichiometric saturation.  This implies
that Al release from resuspension occurs only in bottom waters with relatively
low \chem{Si_{diss}}, that is the northern North Atlantic Ocean and the Arctic
Ocean \citep{middag2009}, while in all other ocean basins the high
\chem{Si_{diss}} in bottom waters prevents such Al release. In fact, within the
Arctic Ocean, indeed in bottom waters, there is an elevated trend of the
dissolved Al/Si ratio as compared to midwaters, i.e.\ an indication of extra
release of Al \citep[their Figures 5,8,13,18]{middag2009}.

A dataset of measurements of \chem{[Al_{diss}]} in the deep and bottom waters
has been used, that is much larger than hitherto available. Nevertheless, the
vertical resolution near the deep ocean seafloor still is modest. Similarly,
the deepest bottom water box of the model extends to 500\,m above the seafloor,
and in some regions the model extent of 5\,000\,m is less than the true full
water column depth. Obviously, the very intriguing sediment source of Al in the
40 to 65{\degree}\,N region would be of great interest for a more detailed
study with high vertical resolution sampling just above the seafloor and
similar high vertical resolution modelling.  Also porewater dynamics may
potentially be important, hence it may be necessary to include a more detailed
parameterisation in future models.  Similarly, other types of particulate Al in
resuspended sediments should be considered in modelling.

Simulations with biological incorporation show that this process is unlikely to
occur proportional to the ambient dissolved \chem{Al}/\chem{Si} concentration
ratio.  The simulations suggest that the relative importance of incorporation
compared to scavenging may be small, because changing the scavenging parameters
or surface dust dissolution cannot compensate for the unrealistic decrease of
dissolved Al in the main thermocline.  This does not imply that incorporation
does not take place, yet perhaps net incorporation is relatively small.

Clearly, more simulations, laboratory experiments and field observations are
needed to answer what the relative amount of incorporation is compared to
scavenging.  When a realistic model of incorporation has been developed, the
next step is to test the effect of Al on the Si cycle which could finally shed
light on how large this effect is for the world ocean and the role of diatoms
in the climate system.

Finally a word of caution. On the one hand, the resulting overall improved
simulation or fit versus the measurements of \chem{[Al_{diss}]} and
\chem{[Si_{diss}]} is another step forward. On the other hand, for such a
complex circulation-biogeochemistry model with so many parameterisations, one
cannot exclude the possibility of other combinations of parameterisations
resulting in a similar or perhaps even better goodness of fit to the
measurements.  In other words, while the chosen processes of Al supply and Al
removal are sensible also in keeping with views in the literature, as is the
ensuing fair simulation, the findings should not be overinterpreted as
conclusive evidence in support of the chosen processes and their
parameterisation.

\begin{appendices}
\section{Sediment resuspension model}         \label{sec:alu_bg:derivation}
Based on observations, \cite{mackin1986} hypothesised that release of Al from
sediment is inhibited by porewater silicic acid (\chem{Si(OH)_4}).  They wrote
the following, using \chem{\{\,\}} for chemical activity and $p$ for -log$_{10}$
as also in pH:
\begin{quotation}
  In general, when stoichiometric saturation (sensu \cite{thorstenson1977})
  exists for an authigenic clay of constant composition in sediment porewater
  having nearly invariant reactive cation concentrations, the following will hold
  \citep{mackin1984dissolved}.
  \begin{equation}
      p\{\chem{Al(OH)_4^-}\} + ap\{\chem{Si(OH)_4}\} + b\,\text{pH} = pK_\text{eq} \;,
      \label{eqn:mackin}
  \end{equation}
  where [$a$ = Si/Al and $b =$ H$^+$/Al are the stoichiometries] of the clay and
  $pK_\text{eq}$ = apparent constant excluding the effects of major cations and
  other potentially reactive cations.  To estimate values of $a$ and $b$ for the
  Amazon shelf sediments from [their] Fig.~2, we applied a regression technique
  which treats $p$\chem{\{Al(OH)_4^-\}}, $p$\chem{\{Si(OH)_4\}} and pH as
  independent variables \citep{mackin1984dissolved}.  The results of this
  treatment give the following:
  \begin{align*}
      1(\pm 0.044) p\chem{\{Al(OH)_4^-\}} &+ 0.828(\pm 0.093)p\chem{\{Si(OH)_4\}} \\
                                   &+ 0.429(\pm 0.070)\text{pH} = p K_\text{eq}   \;,
  \end{align*}
  where $p K_\text{eq} = 13.98(\pm 0.13)$.
\end{quotation}
The chemical activities are in \unit{nM} and \unit{\mu M} for \chem{Al(OH)_4^-}
and \chem{Si(OH)_4}, respectively.  Their use of stoichiometric saturation is
consistent with the mirror image between \chem{[Al(OH)_4^-]} and
\chem{[Si(OH)_4]} as shown in Fig.~\ref{fig:observations}.  Here
\chem{[Al(OH)_4^-]_{bottom}} is small in the Southern Hemisphere and large in
the Northern Hemisphere, while both south of 40{\degree}\,S and north of
40\degree N sediment particles are present in the bottom water
(Fig.~\ref{fig:beam}), strongly suggesting sediment resuspension.  Actually,
just south of 40{\degree}\,N is the largest sediment resuspension, even though
just north of this latitude, \chem{[Al_{diss}]} is mostly elevated near the
sediment.

Eq.~(\ref{eqn:mackin}) can be rewritten to:
\begin{equation}
    \log_{10} \chem{\{Al_{diss}\}}_\text{pore} + a \cdot \log_{10} \chem{\{Si_{diss}\}}_\text{pore} = B   \;,
    \label{eqn:mackin_rewritten}
\end{equation}
where \chem{Al_{diss}} = \chem{Al(OH)_4^-}, \chem{Si_{diss}} = \chem{Si(OH)_4},
and with $B = b \text{pH} - pK_\text{eq}$ being an approximate constant (for
$\text{pH} = 8.1$, $B = -10.5$).  The dissolved entities between the curly braces are
\textit{chemical activities} in the \textit{porewater}, but we need to model
\textit{fluxes} from \textit{resuspended sediment} into \textit{bottom water}.
We will refer to the model layer (of max.~500\,m thick) just above the sediment
as \textit{bottom water}.

Since \chem{[Al_{diss}]} and \chem{[Si_{diss}]} are high in porewater (at least
\chem{[Si_{diss}]} is very high in Southern Ocean porewater), the chemical
activities are not equal to the concentrations.  But they are proportional
where their coefficients of proportionality are the activity coefficients
$\gamma_\text{Al}$ and $\gamma_\text{Si}$ \citep[e.g.][]{kinetics:stone1990}:
\begin{align}
  \begin{aligned}
    \chem{\{Al_{diss}\}}_\text{pore} &= \gamma_\text{Al} \chem{[Al_{diss}]}_\text{pore}   \\
    \chem{\{Si_{diss}\}}_\text{pore} &= \gamma_\text{Si} \chem{[Si_{diss}]}_\text{pore}   \;.
  \end{aligned}
  \label{eqn:activity}
\end{align}

So far we rewrote the equations of \cite{mackin1986} using simple mathematics
and chemistry.  We will now introduce the model.  For this purpose we will
assume that the empirical relationship of \cite{mackin1986}
(Eq.~\eqref{eqn:mackin}), found at the Amazon shelf, is valid everywhere.  This
assumption is defendable if \cite{mackin1986} have not made any extra implicit
assumptions on top of the research of \cite{thorstenson1977}; the latter only
used established thermodynamical relations.

From Eqs~(\ref{eqn:mackin_rewritten}) and (\ref{eqn:activity}) the following
relation can be derived:
\begin{equation}
  \begin{aligned}
    \log_{10} (\gamma_\text{Al} \chem{[Al_{diss}]}_\text{pore}) &= B - a\cdot
    \log_{10} (\gamma_\text{Si} \chem{[Si_{diss}]}_\text{pore}) \\
    \chem{[Al_{diss}]}_\text{pore}/\unit{nM} &= \frac{10^B}{\gamma_\text{Al}} \cdot
    \left(\gamma_\text{Si}\chem{[Si_{diss}]}_\text{pore}/\unit{\mu M}\right)^{-a} \\
                        &= C \cdot (\chem{[Si_{diss}]}_\text{pore}/\unit{\mu M})^{-a} \;,
  \end{aligned}
  \label{eqn:mackin_poreconc}
\end{equation}
where $C = 10^B \gamma_\text{Al}^{-1} \gamma_\text{Si}^{-a} > 0$.

The Al flux $\mathit{\Phi}_\text{sed}$ from the bottom water layer to the
sediment is given by $\mathit{\Phi}_\text{sed} = w_s \cdot \chem{[Al_{ads}]}$,
with \chem{[Al_{ads}]} the bottom water layer concentration of adsorbed Al and
$w_s$ the sedimentation rate.  In line with the sediment resuspension
hypothesis, we assume that the aluminium flux $\mathit{\Phi}_\text{resusp}$
from resuspension and subsequent release of \chem{Al_{diss}} (henceforth
\emph{resuspension flux}) is proportional to $\mathit{\Phi}_\text{sed}$,
converting part of the sedimented Al into dissolved Al in the bottom water
layer:
\begin{equation}
    \mathit{\Phi}_\text{resusp} = \beta \cdot \mathit{\Phi}_\text{sed}
                       = \beta \cdot w_s \cdot \chem{[Al_{ads}]} \;,
    \label{eqn:beta}
\end{equation}
where $\beta \in [0,1]$ is the fraction of resuspended and subsequently
released Al.  In the original resuspension model, $\beta$ is a~constant.  In
the more complex model, the resuspension flux is taken to be proportional to
$\chem{[Al_{diss}]}_\text{pore}$ (as well as \chem{[Al_{ads}]}).  Using
Eq.~\eqref{eqn:mackin_poreconc}, this results in:
\begin{equation}
    \beta \propto \chem{[Al_{diss}]}_\text{pore} /\unit{nM}
          \propto (\chem{[Si_{diss}]}_\text{pore}/\unit{\mu M})^{-a}  \;.
    \label{eqn:bottomprop}
\end{equation}
Furthermore, we assume that bottom water \chem{[Si_{diss}]} is proportional to
porewater \chem{[Si_{diss}]}.  It is a first-order assumption.  Since we lack
the knowledge whether this is reasonable, it would be useful if this assumption
were tested, but that is far outside the scope of this work.  Therefore, we
cannot give an explanation or perform a mechanistic simulation.  Together with
Eqs~(\ref{eqn:beta}) and (\ref{eqn:bottomprop}), this assumption gives for the
change of \chem{[Al_{diss}]} in the bottom layer due to resuspension:
\begin{equation}
  \begin{aligned}
    \frac{\partial\chem{[Al_{diss}]}}{\partial t} \Big|_\text{resusp}
        &= \frac{\mathit{\Phi}_\text{resusp}}{\Delta z_\text{bottom}}
         = \beta \cdot \frac{\mathit{\Phi}_\text{sed}}{\Delta z_\text{bottom}} \\
        &= \beta_0 \left(\frac{\chem{[Si_{diss}]}_\text{bottom}}{\unit{\mu M}}\right)^{-a} \cdot
          \frac{\mathit{\Phi}_\text{sed}}{\Delta z_\text{bottom}} \;,
  \end{aligned}
  \label{eqn:model}
\end{equation}
where $\beta_0$ is a~dimensionless constant and $\Delta z_\text{bottom}$ is the
thickness of the bottom seawater model layer.

Since there can be no more redissolution than the amount of Al that sediments
(assuming steady state and no horizontal remobilisation of sediment), $\beta$
is at most one.  If we want to use the highest possible flux for resuspension
near 45--50{\degree}\,N, we set $\beta = 1$ in that region.  The modelled
bottom water \chem{[Si_{diss}]} near 50{\degree}\,N is $30.3\;\unit{\mu M}$.
Hence, the proportionality constant is $\beta_0 = 30.3^{0.828} = 16.85$.  Since
bottom \chem{[Si_{diss}]} farther north is lower, $\beta$ would there be higher
than one.  Since the model assumes that only recently sedimented
\chem{Al_{ads}} is resuspended, it would not be physically correct if $\beta >
1$.  Therefore $\beta$ is constrained and hence given by:
\begin{equation}
  \beta = \text{min}(16.85 \cdot (\chem{[Si_{diss}]} / \unit{\mu M})^{-0.828}, 1)\;,
\end{equation}
which is to be plugged into the concentration change:
\begin{equation}
    \frac{\partial\chem{[Al_{diss}]}}{\partial t} \Big|_\text{resusp}
        = \beta \cdot \frac{\mathit{\Phi}_\text{sed}}{\Delta z_\text{bottom}} \;.
    \label{eqn:resusp_appendix}
\end{equation}
This equation is identical to Eq.~(\ref{eqn:resusp}) in the main text, hence
this result is what was to be demonstrated.

As a~sanity check we substitute the minimum (13.4\;\unit{\mu M}) and the
maximum (148.6\;\unit{\mu M}) bottom \chem{[Si_{diss}]} into this equation:
\begin{align*}
    \beta_\text{max} &= \text{min}(16.85 \cdot 13.4^{-0.828},1)
                      = \text{min}(1.96,1) = 1        \\
    \beta_\text{min} &= \text{min}(16.85 \cdot 148.6^{-0.828},1) = 0.27
\end{align*}
This means that since bottom \chem{[Si_{diss}]} is decreasing with latitude, we
expect 100\,{\%} redissolution anywhere north of 45{\degree}\,N (it is set up
like that).  About 25\,{\%} of the sedimenting \chem{Al_{ads}} is dissolved in
the Southern Ocean (high \chem{[Si_{diss}]}).
\end{appendices}

\Supplementary{zip}

\begin{acknowledgements}
The authors are grateful to those who have been proved useful in discussion,
among which Justus van Beusekom, Tom Remenyi, Micha Rijkenberg, Wilco Hazeleger
and our colleagues from the IMAU institute.  We also want to thank Christoph
Heinze, Phoebe Lam and Andreas Schmittner for providing their data.  The
authors wish to acknowledge the use of the free software visualisation and
analysis programs \href{http://ferret.pmel.noaa.gov/Ferret/}{Ferret} and
\href{http://www.r-project.org/}{R}.  Other noteworthy libre software used is
the \href{http://www.gnu.org/}{GNU{\ }Operating{\ }System} and
\href{https://code.zmaw.de/projects/cdo}{Climate{\ }Data{\ }Operators}.  Useful
tips for statistical analysis were provided by Ronald van Haren, Carlo
Lacagnina and Paul Hiemstra.  We would also like to thank the three reviewers
(two anonymous and Justus van Beusekom) for their suggestions for improving the
manuscript.  This research is funded by the Netherlands Organisation for
Scientific Research (NWO), grant Nr.~839.08.414, part of the ZKO programme.
\end{acknowledgements}

\DeclareRobustCommand{\DutchName}[4]{#1, #3~#4}
\bibliographystyle{copernicus}
\bibliography{articles,books}
\end{document}